

\def\hb{\hfil\break}

\def\Ref#1{Ref.\ #1}  
\def\ref#1{Ref.\ #1}  
\def\head#1{                    
  \goodbreak\vskip 0.5truein    
  {\immediate\write16{#1}
   \centerline{\uppercase{#1}}}
   \nobreak\vskip 0.25truein\nobreak}
\def\subhead#1{                 
  \vskip 0.25truein             
  \centerline{#1}
  \nobreak\vskip 0.25truein\nobreak}
\message
{SCRI JNL.TEX version 0.92 as of 13-JUL-1989. Report bugs and problems to ADK.}

\catcode`@=11
\expandafter\ifx\csname inp@t\endcsname\relax\let\inp@t=\input
\def\input#1 {\expandafter\ifx\csname #1IsLoaded\endcsname\relax
\inp@t#1%
\expandafter\def\csname #1IsLoaded\endcsname{(#1 was previously loaded)}
\else\message{\csname #1IsLoaded\endcsname}\fi}\fi
\catcode`@=12



\font\twelverm=cmr12    \font\twelvei=cmmi12
\font\twelvesy=cmsy10 scaled 1200   \font\twelveex=cmex10 scaled 1200
\font\twelvebf=cmbx12   \font\twelvesl=cmsl12
\font\twelvett=cmtt12   \font\twelveit=cmti12
\font\twelvesc=cmcsc10 scaled 1200  \font\twelvesf=cmssdc10 scaled 1200
\skewchar\twelvei='177   \skewchar\twelvesy='60


\def\twelvepoint{\normalbaselineskip=12.4pt plus 0.1pt minus 0.1pt
  \abovedisplayskip 12.4pt plus 3pt minus 9pt
  \belowdisplayskip 12.4pt plus 3pt minus 9pt
  \abovedisplayshortskip 0pt plus 3pt
  \belowdisplayshortskip 7.2pt plus 3pt minus 4pt
  \smallskipamount=3.6pt plus1.2pt minus1.2pt
  \medskipamount=7.2pt plus2.4pt minus2.4pt
  \bigskipamount=14.4pt plus4.8pt minus4.8pt
  \def\rm{\fam0\twelverm}          \def\it{\fam\itfam\twelveit}%
  \def\sl{\fam\slfam\twelvesl}     \def\bf{\fam\bffam\twelvebf}%
  \def\mit{\fam 1}                 \def\cal{\fam 2}%
  \def\sc{\twelvesc}		   \def\tt{\twelvett}
  \def\sf{\twelvesf}
  \textfont0=\twelverm   \scriptfont0=\tenrm   \scriptscriptfont0=\sevenrm
  \textfont1=\twelvei    \scriptfont1=\teni    \scriptscriptfont1=\seveni
  \textfont2=\twelvesy   \scriptfont2=\tensy   \scriptscriptfont2=\sevensy
  \textfont3=\twelveex   \scriptfont3=\twelveex  \scriptscriptfont3=\twelveex
  \textfont\itfam=\twelveit
  \textfont\slfam=\twelvesl
  \textfont\bffam=\twelvebf \scriptfont\bffam=\tenbf
  \scriptscriptfont\bffam=\sevenbf
  \normalbaselines\rm}



\def\beginlinemode{\endmode
  \begingroup\parskip=0pt \obeylines\def\\{\par}\def\endmode{\par\endgroup}}
\def\beginparmode{\endmode
  \begingroup \def\endmode{\par\endgroup}}
\let\endmode=\par
{\obeylines\gdef\
{}}
\def\singlespace{\baselineskip=\normalbaselineskip}

\def\oneandahalfspace{\baselineskip=\normalbaselineskip
  \multiply\baselineskip by 3 \divide\baselineskip by 2}
\def\doublespace{\baselineskip=\normalbaselineskip \multiply\baselineskip by 2}

\newcount\firstpageno
\firstpageno=2
\footline={\ifnum\pageno<\firstpageno{\hfil}\else{\hfil\twelverm\folio\hfil}\fi}
\def\toppageno{\global\footline={\hfil}\global\headline
  ={\ifnum\pageno<\firstpageno{\hfil}\else{\hfil\twelverm\folio\hfil}\fi}}
\let\rawfootnote=\footnote		
\def\footnote#1#2{{\rm\singlespace\parindent=0pt\parskip=0pt
  \rawfootnote{#1}{#2\hfill\vrule height 0pt depth 6pt width 0pt}}}
\def\raggedcenter{\leftskip=4em plus 12em \rightskip=\leftskip
  \parindent=0pt \parfillskip=0pt \spaceskip=.3333em \xspaceskip=.5em
  \pretolerance=9999 \tolerance=9999
  \hyphenpenalty=9999 \exhyphenpenalty=9999 }
\def\dateline{\rightline{\ifcase\month\or
  January\or February\or March\or April\or May\or June\or
  July\or August\or September\or October\or November\or December\fi
  \space\number\year}}
\def\received{\vskip 3pt plus 0.2fill
 \centerline{\sl (Received\space\ifcase\month\or
  January\or February\or March\or April\or May\or June\or
  July\or August\or September\or October\or November\or December\fi
  \qquad, \number\year)}}


\hsize=6.5truein
\vsize=8.9truein
\parskip=\medskipamount
\def\\{\cr}
\twelvepoint		
\doublespace		
\overfullrule=0pt	




\def\title			
  {\null\vskip 3pt plus 0.2fill
   \beginlinemode \doublespace \raggedcenter \bf}

\def\author			
  {\vskip 3pt plus 0.2fill \beginlinemode
   \singlespace \raggedcenter\sc}

\def\affil			
  {\vskip 3pt plus 0.1fill \beginlinemode
   \oneandahalfspace \raggedcenter \sl}

\def\abstract			
  {\vskip 3pt plus 0.3fill \beginparmode
   \oneandahalfspace ABSTRACT: }

\def\endtitlepage		
  {\endpage			
   \body}
\let\endtopmatter=\endtitlepage

\def\body			
  {\beginparmode}		

\def\head#1{			
  \goodbreak\vskip 0.5truein	
  {\immediate\write16{#1}
   \raggedcenter \uppercase{#1}\par}
   \nobreak\vskip 0.25truein\nobreak}

\def\subhead#1{			
  \vskip 0.25truein		
  {\raggedcenter {#1} \par}
   \nobreak\vskip 0.25truein\nobreak}

\def\beginitems{
\par\medskip\bgroup\def\i##1 {\item{##1}}\def\ii##1 {\itemitem{##1}}
\leftskip=36pt\parskip=0pt}
\def\enditems{\par\egroup}

\def\beneathrel#1\under#2{\mathrel{\mathop{#2}\limits_{#1}}}

\def\refto#1{$^{#1}$}		

\def\references			
  {\head{References}		
   \beginparmode
   \frenchspacing \parindent=0pt \leftskip=1truecm
   \parskip=8pt plus 3pt \everypar{\hangindent=\parindent}}

\gdef\refis#1{\item{#1.\ }}			

\gdef\journal#1, #2, #3, 1#4#5#6{		
    {\sl #1~}{\bf #2}, #3 (1#4#5#6)}		

\def\endreferences{\body}

\def\figurecaptions		
  {\endpage
   \beginparmode
   \head{Figure Captions}
}

\def\endfigurecaptions{\body}

\def\endpage			
  {\vfill\eject}

\def\endpaper			
  {\endmode\vfill\supereject}

\def\endit
  {\endpaper\end}


\def\heading				
  {\vskip 0.5truein plus 0.1truein	
   \beginparmode \def\\{\par} \parskip=0pt \singlespace \raggedcenter}

\def\subheading				
  {\vskip 0.25truein plus 0.1truein	
   \beginlinemode \singlespace \parskip=0pt \def\\{\par}\raggedcenter}

\def\tag#1$${\eqno(#1)$$}

\def\align#1$${\eqalign{#1}$$}

\def\aligntag#1$${\gdef\tag##1\\{&(##1)\cr}\eqalignno{#1\\}$$
  \gdef\tag##1$${\eqno(##1)$$}}

\def\endaligntag{}

\def\overset #1\to#2{{\mathop{#2}\limits^{#1}}}
\def\underset#1\to#2{{\let\next=#1\mathpalette\undersetpalette#2}}
\def\undersetpalette#1#2{\vtop{\baselineskip0pt
\ialign{$\mathsurround=0pt #1\hfil##\hfil$\crcr#2\crcr\next\crcr}}}


\def\ref#1{Ref.~#1}			
\def\Ref#1{Ref.~#1}			
\def\[#1]{[\cite{#1}]}
\def\cite#1{{#1}}
\def\(#1){(\call{#1})}
\def\call#1{{#1}}
\def\taghead#1{}
\def\frac#1#2{{#1 \over #2}}

\def\12{{1\over2}}
\def\eg{{\it e.g.,\ }}

\def\ie{{\it i.e.,\ }}

\def\sla{\raise.15ex\hbox{$/$}\kern-.57em}
\def\leaderfill{\leaders\hbox to 1em{\hss.\hss}\hfill}
\def\twiddle{\lower.9ex\rlap{$\kern-.1em\scriptstyle\sim$}}
\def\bigtwiddle{\lower1.ex\rlap{$\sim$}}
\def\gtwid{\mathrel{\raise.3ex\hbox{$>$\kern-.75em\lower1ex\hbox{$\sim$}}}}
\def\ltwid{\mathrel{\raise.3ex\hbox{$<$\kern-.75em\lower1ex\hbox{$\sim$}}}}
\def\square{\kern1pt\vbox{\hrule height 1.2pt\hbox{\vrule width 1.2pt\hskip 3pt
   \vbox{\vskip 6pt}\hskip 3pt\vrule width 0.6pt}\hrule height 0.6pt}\kern1pt}
\def\tdot#1{\mathord{\mathop{#1}\limits^{\kern2pt\ldots}}}

\def\pmb#1{\setbox0=\hbox{#1}%
  \kern-.025em\copy0\kern-\wd0
  \kern  .05em\copy0\kern-\wd0
  \kern-.025em\raise.0433em\box0 }

\catcode`@=11
\newcount\r@fcount \r@fcount=0
\newcount\r@fcurr
\immediate\newwrite\reffile
\newif\ifr@ffile\r@ffilefalse
\def\w@rnwrite#1{\ifr@ffile\immediate\write\reffile{#1}\fi\message{#1}}

\def\writer@f#1>>{}
\def\referencefile{
  \r@ffiletrue\immediate\openout\reffile=\jobname.ref%
  \def\writer@f##1>>{\ifr@ffile\immediate\write\reffile%
    {\noexpand\refis{##1} = \csname r@fnum##1\endcsname = %
     \expandafter\expandafter\expandafter\strip@t\expandafter%
     \meaning\csname r@ftext\csname r@fnum##1\endcsname\endcsname}\fi}%
  \def\strip@t##1>>{}}

\def\citeall#1{\xdef#1##1{#1{\noexpand\cite{##1}}}}
\def\cite#1{\each@rg\citer@nge{#1}}	

\def\each@rg#1#2{{\let\thecsname=#1\expandafter\first@rg#2,\end,}}
\def\first@rg#1,{\thecsname{#1}\apply@rg}	
\def\apply@rg#1,{\ifx\end#1\let\next=\relax
\else,\thecsname{#1}\let\next=\apply@rg\fi\next}

\def\citer@nge#1{\citedor@nge#1-\end-}	
\def\citer@ngeat#1\end-{#1}
\def\citedor@nge#1-#2-{\ifx\end#2\r@featspace#1 
  \else\citel@@p{#1}{#2}\citer@ngeat\fi}	
\def\citel@@p#1#2{\ifnum#1>#2{\errmessage{Reference range #1-#2\space is bad.}%
    \errhelp{If you cite a series of references by the notation M-N, then M and
    N must be integers, and N must be greater than or equal to M.}}\else%
 {\count0=#1\count1=#2\advance\count1 by1\relax\expandafter\r@fcite\the\count0,%
  \loop\advance\count0 by1\relax
    \ifnum\count0<\count1,\expandafter\r@fcite\the\count0,%
  \repeat}\fi}

\def\r@featspace#1#2 {\r@fcite#1#2,}	
\def\r@fcite#1,{\ifuncit@d{#1}
    \newr@f{#1}%
    \expandafter\gdef\csname r@ftext\number\r@fcount\endcsname%
                     {\message{Reference #1 to be supplied.}%
                      \writer@f#1>>#1 to be supplied.\par}%
 \fi%
 \csname r@fnum#1\endcsname}
\def\ifuncit@d#1{\expandafter\ifx\csname r@fnum#1\endcsname\relax}%
\def\newr@f#1{\global\advance\r@fcount by1%
    \expandafter\xdef\csname r@fnum#1\endcsname{\number\r@fcount}}

\let\r@fis=\refis			
\def\refis#1#2#3\par{\ifuncit@d{#1}
   \newr@f{#1}%
   \w@rnwrite{Reference #1=\number\r@fcount\space is not cited up to now.}\fi%
  \expandafter\gdef\csname r@ftext\csname r@fnum#1\endcsname\endcsname%
  {\writer@f#1>>#2#3\par}}

\def\ignoreuncited{
   \def\refis##1##2##3\par{\ifuncit@d{##1}%
     \else\expandafter\gdef\csname r@ftext\csname r@fnum##1\endcsname\endcsname%
     {\writer@f##1>>##2##3\par}\fi}}

\def\r@ferr{\endreferences\errmessage{I was expecting to see
\noexpand\endreferences before now;  I have inserted it here.}}
\let\r@ferences=\references
\def\references{\r@ferences\def\endmode{\r@ferr\par\endgroup}}

\let\endr@ferences=\endreferences
\def\endreferences{\r@fcurr=0
  {\loop\ifnum\r@fcurr<\r@fcount
    \advance\r@fcurr by 1\relax\expandafter\r@fis\expandafter{\number\r@fcurr}%
    \csname r@ftext\number\r@fcurr\endcsname%
  \repeat}\gdef\r@ferr{}\endr@ferences}


\let\r@fend=\endpaper\gdef\endpaper{\ifr@ffile
\immediate\write16{Cross References written on []\jobname.REF.}\fi\r@fend}

\catcode`@=12

\citeall\refto		
\citeall\ref		%
\citeall\Ref		%

\catcode`@=11
\newcount\tagnumber\tagnumber=0

\immediate\newwrite\eqnfile
\newif\if@qnfile\@qnfilefalse
\def\write@qn#1{}
\def\writenew@qn#1{}
\def\w@rnwrite#1{\write@qn{#1}\message{#1}}
\def\@rrwrite#1{\write@qn{#1}\errmessage{#1}}

\def\taghead#1{\gdef\t@ghead{#1}\global\tagnumber=0}
\def\t@ghead{}

\expandafter\def\csname @qnnum-3\endcsname
  {{\t@ghead\advance\tagnumber by -3\relax\number\tagnumber}}
\expandafter\def\csname @qnnum-2\endcsname
  {{\t@ghead\advance\tagnumber by -2\relax\number\tagnumber}}
\expandafter\def\csname @qnnum-1\endcsname
  {{\t@ghead\advance\tagnumber by -1\relax\number\tagnumber}}
\expandafter\def\csname @qnnum0\endcsname
  {\t@ghead\number\tagnumber}
\expandafter\def\csname @qnnum+1\endcsname
  {{\t@ghead\advance\tagnumber by 1\relax\number\tagnumber}}
\expandafter\def\csname @qnnum+2\endcsname
  {{\t@ghead\advance\tagnumber by 2\relax\number\tagnumber}}
\expandafter\def\csname @qnnum+3\endcsname
  {{\t@ghead\advance\tagnumber by 3\relax\number\tagnumber}}

\def\equationfile{%
  \@qnfiletrue\immediate\openout\eqnfile=\jobname.eqn%
  \def\write@qn##1{\if@qnfile\immediate\write\eqnfile{##1}\fi}
  \def\writenew@qn##1{\if@qnfile\immediate\write\eqnfile
    {\noexpand\tag{##1} = (\t@ghead\number\tagnumber)}\fi}
}

\def\callall#1{\xdef#1##1{#1{\noexpand\call{##1}}}}
\def\call#1{\each@rg\callr@nge{#1}}

\def\each@rg#1#2{{\let\thecsname=#1\expandafter\first@rg#2,\end,}}
\def\first@rg#1,{\thecsname{#1}\apply@rg}
\def\apply@rg#1,{\ifx\end#1\let\next=\relax%
\else,\thecsname{#1}\let\next=\apply@rg\fi\next}

\def\callr@nge#1{\calldor@nge#1-\end-}
\def\callr@ngeat#1\end-{#1}
\def\calldor@nge#1-#2-{\ifx\end#2\@qneatspace#1 %
  \else\calll@@p{#1}{#2}\callr@ngeat\fi}
\def\calll@@p#1#2{\ifnum#1>#2{\@rrwrite{Equation range #1-#2\space is bad.}
\errhelp{If you call a series of equations by the notation M-N, then M and
N must be integers, and N must be greater than or equal to M.}}\else%
 {\count0=#1\count1=#2\advance\count1 by1\relax\expandafter\@qncall\the\count0,%
  \loop\advance\count0 by1\relax%
    \ifnum\count0<\count1,\expandafter\@qncall\the\count0,%
  \repeat}\fi}

\def\@qneatspace#1#2 {\@qncall#1#2,}
\def\@qncall#1,{\ifunc@lled{#1}{\def\next{#1}\ifx\next\empty\else
  \w@rnwrite{Equation number \noexpand\(>>#1<<) has not been defined yet.}
  >>#1<<\fi}\else\csname @qnnum#1\endcsname\fi}

\let\eqnono=\eqno
\def\eqno(#1){\tag#1}
\def\tag#1$${\eqnono(\displayt@g#1 )$$}

\def\aligntag#1\endaligntag
  $${\gdef\tag##1\\{&(##1 )\cr}\eqalignno{#1\\}$$
  \gdef\tag##1$${\eqnono(\displayt@g##1 )$$}}

\def\eqalignno#1{\displ@y \tabskip\centering
  \halign to\displaywidth{\hfil$\displaystyle{##}$\tabskip\z@skip
    &$\displaystyle{{}##}$\hfil\tabskip\centering
    &\llap{$\displayt@gpar##$}\tabskip\z@skip\crcr
    #1\crcr}}

\def\displayt@gpar(#1){(\displayt@g#1 )}

\def\displayt@g#1 {\rm\ifunc@lled{#1}\global\advance\tagnumber by1
        {\def\next{#1}\ifx\next\empty\else\expandafter
        \xdef\csname @qnnum#1\endcsname{\t@ghead\number\tagnumber}\fi}%
  \writenew@qn{#1}\t@ghead\number\tagnumber\else
        {\edef\next{\t@ghead\number\tagnumber}%
        \expandafter\ifx\csname @qnnum#1\endcsname\next\else
        \w@rnwrite{Equation \noexpand\tag{#1} is a duplicate number.}\fi}%
  \csname @qnnum#1\endcsname\fi}

\def\ifunc@lled#1{\expandafter\ifx\csname @qnnum#1\endcsname\relax}

\let\@qnend=\end\gdef\end{\if@qnfile
\immediate\write16{Equation numbers written on []\jobname.EQN.}\fi\@qnend}

\catcode`@=12


%
\newbox\hdbox%
\newcount\hdrows%
\newcount\multispancount%
\newcount\ncase%
\newcount\ncols
\newcount\nrows%
\newcount\nspan%
\newcount\ntemp%
\newdimen\hdsize%
\newdimen\newhdsize%
\newdimen\parasize%
\newdimen\spreadwidth%
\newdimen\thicksize%
\newdimen\thinsize%
\newdimen\tablewidth%
\newif\ifcentertables%
\newif\ifendsize%
\newif\iffirstrow%
\newif\iftableinfo%
\newtoks\dbt%
\newtoks\hdtks%
\newtoks\savetks%
\newtoks\tableLETtokens%
\newtoks\tabletokens%
\newtoks\widthspec%
%
%
\immediate\write15{%
CP SMSG GJMSINK TEXTABLE --> TABLE MACROS V. 851121 JOB = \jobname%
}%
%
%
\tableinfotrue%
\catcode`\@=11
%
%
\def\tstrut{\vrule height3.1ex depth1.2ex width0pt}%
\def\and{\char`\&}
\def\tablerule{\noalign{\hrule height\thinsize depth0pt}}%
\thicksize=1.5pt
\thinsize=0.6pt
\def\thickrule{\noalign{\hrule height\thicksize depth0pt}}%
\def\ctr#1{\hfil\ #1\hfil}%
%
%
%
%
\tablewidth=-\maxdimen%
\spreadwidth=-\maxdimen%
\def\tabskipglue{0pt plus 1fil minus 1fil}%
%
%
\centertablestrue%
\def\noncenteredtables{%
   \centertablesfalse%
}%
%
%
\parasize=4in%
\gdef\ARGS{########}
\gdef\headerARGS{####}
\def\@mpersand{&}
{\catcode`\|=13
\gdef\letbarzero{\let|0}
\gdef\letbartab{\def|{&&}}%
\gdef\letvbbar{\let\vb|}%
}
{\catcode`\&=4
\def\ampskip{&\omit\hfil&}
\catcode`\&=13
\let&0
\xdef\letampskip{\def&{\ampskip}}%
\gdef\letnovbamp{\let\novb&\let\tab&}
}
\def\begintable{
   \begingroup%
   \catcode`\|=13\letbartab\letvbbar%
   \catcode`\&=13\letampskip\letnovbamp%
   \def\multispan##1{
      \omit \mscount##1%
      \multiply\mscount\tw@\advance\mscount\m@ne%
      \loop\ifnum\mscount>\@ne \sp@n\repeat%
   }
   \def\|{%
      &\omit\widevline&%
   }%
   \ruledtable
}
\long\def\ruledtable#1\endtable{%
%
%
%
   \offinterlineskip
   \tabskip 0pt
   \def\widevline{\vrule width\thicksize}
   \def\endrow{\@mpersand\omit\hfil\crnorm\@mpersand}%
   \def\crthick{\@mpersand\crnorm\thickrule\@mpersand}%
   \def\crthickneg##1{\@mpersand\crnorm\thickrule
          \noalign{{\skip0=##1\vskip-\skip0}}\@mpersand}%
   \def\crnorule{\@mpersand\crnorm\@mpersand}%
   \def\crnoruleneg##1{\@mpersand\crnorm
          \noalign{{\skip0=##1\vskip-\skip0}}\@mpersand}%
   \let\nr=\crnorule
   \def\endtable{\@mpersand\crnorm\thickrule}%
   \let\crnorm=\cr
%
%
   \edef\cr{\@mpersand\crnorm\tablerule\@mpersand}%
   \def\crneg##1{\@mpersand\crnorm\tablerule
          \noalign{{\skip0=##1\vskip-\skip0}}\@mpersand}%
   \let\ctneg=\crthickneg
   \let\nrneg=\crnoruleneg
   \the\tableLETtokens
%
%
   \tabletokens={&#1}
%
%
   \countROWS\tabletokens\into\nrows%
   \countCOLS\tabletokens\into\ncols%
%
%
   \advance\ncols by -1%
   \divide\ncols by 2%
   \advance\nrows by 1%
%
%
   \iftableinfo %
      \immediate\write16{[Nrows=\the\nrows, Ncols=\the\ncols]}%
   \fi%
%
%
   \ifcentertables
      \ifhmode \par\fi
      \line{
      \hss
   \else %
      \hbox{%
   \fi
      \vbox{%
         \makePREAMBLE{\the\ncols}
         \edef\next{\preamble}
         \let\preamble=\next
         \makeTABLE{\preamble}{\tabletokens}
      }
      \ifcentertables \hss}\else }\fi
   \endgroup
   \tablewidth=-\maxdimen
   \spreadwidth=-\maxdimen
}
\def\makeTABLE#1#2{
   {
   \let\ifmath0
   \let\header0
   \let\multispan0
%
%
   \ncase=0%
   \ifdim\tablewidth>-\maxdimen \ncase=1\fi%
   \ifdim\spreadwidth>-\maxdimen \ncase=2\fi%
   \relax
%
   \ifcase\ncase %
      \widthspec={}%
   \or %
      \widthspec=\expandafter{\expandafter t\expandafter o%
                 \the\tablewidth}%
   \else %
      \widthspec=\expandafter{\expandafter s\expandafter p\expandafter r%
                 \expandafter e\expandafter a\expandafter d%
                 \the\spreadwidth}%
   \fi %
   \xdef\next{
      \halign\the\widthspec{%
      #1
      \noalign{\hrule height\thicksize depth0pt}
      \the#2\endtable
%
      }
   }
   }
   \next
}
\def\makePREAMBLE#1{
   \ncols=#1
   \begingroup
   \let\ARGS=0
   \edef\xtp{\widevline\ARGS\tabskip\tabskipglue%
   &\ctr{\ARGS}\tstrut}
   \advance\ncols by -1
   \loop
      \ifnum\ncols>0 %
      \advance\ncols by -1%
      \edef\xtp{\xtp&\vrule width\thinsize\ARGS&\ctr{\ARGS}}%
   \repeat
   \xdef\preamble{\xtp&\widevline\ARGS\tabskip0pt%
   \crnorm}
   \endgroup
}
\def\countROWS#1\into#2{
   \let\countREGISTER=#2%
   \countREGISTER=0%
   \expandafter\ROWcount\the#1\endcount%
}%
\def\ROWcount{%
   \afterassignment\subROWcount\let\next= %
}%
\def\subROWcount{%
   \ifx\next\endcount %
      \let\next=\relax%
   \else%
      \ncase=0%
      \ifx\next\cr %
         \global\advance\countREGISTER by 1%
         \ncase=0%
      \fi%
      \ifx\next\endrow %
         \global\advance\countREGISTER by 1%
         \ncase=0%
      \fi%
      \ifx\next\crthick %
         \global\advance\countREGISTER by 1%
         \ncase=0%
      \fi%
      \ifx\next\crnorule %
         \global\advance\countREGISTER by 1%
         \ncase=0%
      \fi%
      \ifx\next\crthickneg %
         \global\advance\countREGISTER by 1%
         \ncase=0%
      \fi%
      \ifx\next\crnoruleneg %
         \global\advance\countREGISTER by 1%
         \ncase=0%
      \fi%
      \ifx\next\crneg %
         \global\advance\countREGISTER by 1%
         \ncase=0%
      \fi%
      \ifx\next\header %
         \ncase=1%
      \fi%
      \relax%
      \ifcase\ncase %
         \let\next\ROWcount%
      \or %
         \let\next\argROWskip%
      \else %
      \fi%
   \fi%
   \next%
}
\def\counthdROWS#1\into#2{%
\dvr{10}%
   \let\countREGISTER=#2%
   \countREGISTER=0%
\dvr{11}%
\dvr{13}%
   \expandafter\hdROWcount\the#1\endcount%
\dvr{12}%
}%
\def\hdROWcount{%
   \afterassignment\subhdROWcount\let\next= %
}%
\def\subhdROWcount{%
   \ifx\next\endcount %
      \let\next=\relax%
   \else%
      \ncase=0%
      \ifx\next\cr %
         \global\advance\countREGISTER by 1%
         \ncase=0%
      \fi%
      \ifx\next\endrow %
         \global\advance\countREGISTER by 1%
         \ncase=0%
      \fi%
      \ifx\next\crthick %
         \global\advance\countREGISTER by 1%
         \ncase=0%
      \fi%
      \ifx\next\crnorule %
         \global\advance\countREGISTER by 1%
         \ncase=0%
      \fi%
      \ifx\next\header %
         \ncase=1%
      \fi%
\relax%
      \ifcase\ncase %
         \let\next\hdROWcount%
      \or%
         \let\next\arghdROWskip%
      \else %
      \fi%
   \fi%
   \next%
}%
{\catcode`\|=13\letbartab
\gdef\countCOLS#1\into#2{%
   \let\countREGISTER=#2%
   \global\countREGISTER=0%
   \global\multispancount=0%
   \global\firstrowtrue
   \expandafter\COLcount\the#1\endcount%
   \global\advance\countREGISTER by 3%
   \global\advance\countREGISTER by -\multispancount
}%
\gdef\COLcount{%
   \afterassignment\subCOLcount\let\next= %
}%
{\catcode`\&=13%
\gdef\subCOLcount{%
   \ifx\next\endcount %
      \let\next=\relax%
   \else%
      \ncase=0%
      \iffirstrow
         \ifx\next& %
            \global\advance\countREGISTER by 2%
            \ncase=0%
         \fi%
         \ifx\next\span %
            \global\advance\countREGISTER by 1%
            \ncase=0%
         \fi%
         \ifx\next| %
            \global\advance\countREGISTER by 2%
            \ncase=0%
         \fi
         \ifx\next\|
            \global\advance\countREGISTER by 2%
            \ncase=0%
         \fi
         \ifx\next\multispan
            \ncase=1%
            \global\advance\multispancount by 1%
         \fi
         \ifx\next\header
            \ncase=2%
         \fi
         \ifx\next\cr       \global\firstrowfalse \fi
         \ifx\next\endrow   \global\firstrowfalse \fi
         \ifx\next\crthick  \global\firstrowfalse \fi
         \ifx\next\crnorule \global\firstrowfalse \fi
         \ifx\next\crnoruleneg \global\firstrowfalse \fi
         \ifx\next\crthickneg  \global\firstrowfalse \fi
         \ifx\next\crneg       \global\firstrowfalse \fi
      \fi
\relax
      \ifcase\ncase %
         \let\next\COLcount%
      \or %
         \let\next\spancount%
      \or %
         \let\next\argCOLskip%
      \else %
      \fi %
   \fi%
   \next%
}%
\gdef\argROWskip#1{%
   \let\next\ROWcount \next%
}
\gdef\arghdROWskip#1{%
   \let\next\ROWcount \next%
}
\gdef\argCOLskip#1{%
   \let\next\COLcount \next%
}
}
}
\def\spancount#1{
   \nspan=#1\multiply\nspan by 2\advance\nspan by -1%
   \global\advance \countREGISTER by \nspan
   \let\next\COLcount \next}%
\def\dvr#1{\relax}%
\def\header#1{%
\dvr{1}{\let\cr=\@mpersand%
\hdtks={#1}%
\counthdROWS\hdtks\into\hdrows%
\advance\hdrows by 1%
\ifnum\hdrows=0 \hdrows=1 \fi%
\dvr{5}\makehdPREAMBLE{\the\hdrows}%
\dvr{6}\getHDdimen{#1}%
{\parindent=0pt\hsize=\hdsize{\let\ifmath0%
\xdef\next{\valign{\headerpreamble #1\crnorm}}}\dvr{7}\next\dvr{8}%
}%
}\dvr{2}}
\def\makehdPREAMBLE#1{
\dvr{3}%
\hdrows=#1
{
\let\headerARGS=0%
\let\cr=\crnorm%
\edef\xtp{\vfil\hfil\hbox{\headerARGS}\hfil\vfil}%
\advance\hdrows by -1
\loop
\ifnum\hdrows>0%
\advance\hdrows by -1%
\edef\xtp{\xtp&\vfil\hfil\hbox{\headerARGS}\hfil\vfil}%
\repeat%
\xdef\headerpreamble{\xtp\crcr}%
}
\dvr{4}}
\def\getHDdimen#1{%
\hdsize=0pt%
\getsize#1\cr\end\cr%
}
\def\getsize#1\cr{%
\endsizefalse\savetks={#1}%
\expandafter\lookend\the\savetks\cr%
\relax \ifendsize \let\next\relax \else%
\setbox\hdbox=\hbox{#1}\newhdsize=1.0\wd\hdbox%
\ifdim\newhdsize>\hdsize \hdsize=\newhdsize \fi%
\let\next\getsize \fi%
\next%
}%
\def\lookend{\afterassignment\sublookend\let\looknext= }%
\def\sublookend{\relax%
\ifx\looknext\cr %
\let\looknext\relax \else %
   \relax
   \ifx\looknext\end \global\endsizetrue \fi%
   \let\looknext=\lookend%
    \fi \looknext%
}%
%
%
\def\tablelet#1{%
   \tableLETtokens=\expandafter{\the\tableLETtokens #1}%
}%
\catcode`\@=12
%

\def\tw#1{ \tilde #1 }            
\def\oh{{\scriptstyle {1\over2}}}
\def\oq{{\scriptstyle {1\over4}}}
\rightline{{\rm November 1989}}
\title
\centerline{Phase Space Wannier Functions in Electronic %
                Structure Calculations}

\author
D.\ J.\ Sullivan,$^a$ J.\ J.\ Rehr,$^b$ J.\ W.\ Wilkins,$^a$ %
and K.\ G.\ Wilson$^a$

\affil
$^a$Department of Physics, Ohio State University, Columbus, OH 43210 %

and %

$^b$Department of Physics, University of Washington, Seattle, WA 98195 %

\abstract
We consider the applicability of ``phase space Wannier functions'' to
electronic structure calculations.  These generalized Wannier functions
are analogous to localized plane waves and constitute a complete,
orthonormal set which is exponentially localized both in position and
momentum space.  Their properties are described and an illustrative
application to bound states in one
dimension is presented.  Criteria for choosing basis set size and
lattice constant are discussed based on semi-classical, phase space
considerations.  Convergence of the ground state energy with respect to
basis size is evaluated.  Comparison with plane-waves basis sets
indicates that the number of phase space Wannier functions
needed for convergence can be
significantly smaller in three dimensions.

\bigskip
\noindent{PACS: 71.10.+x, 71.50.+t}
\endtopmatter

\head{I. Introduction} \taghead{1.}

This paper deals with a new set of generalized Wannier
functions\refto{Wilson} (WF) which we argue has many advantages as a
basis for large scale electronic structure calculations.  The term
Wannier functions refers to a complete and orthonormal set of {\it local\/}
basis functions for representations of scalar wave functions.  An example is
the set of WF which span the same function space as the Bloch functions
for a given energy band, or set of energy bands, in a
crystal.\refto{Kohn} Though they are often useful as a theoretical tool,
energy band WF have seldom been practical in quantitative electronic
structure calculations.  There are several reasons for this: One
difficulty is the numerical complexity of their construction; they are
not the solutions of any simple differential equation and often must be
constructed from non-orthogonal trial functions.  Second, the advantage
of their mutual orthogonality is offset by the existence of long range
oscillatory tails which are necessary to ensure their orthogonality.
Band WF are, however, exponentially localized\refto{Kohn} provided there
is a finite band-gap, and good localization is an essential requirement
in quantitative calculations.  One can define local Wannier functions
which span the same function space as the plane waves in a given
Brillouin zone, but such WF decay only algebraically and are of little
practical use.

``Generic'' basis functions, \ie functions with similar analytical
structure, are particularly advantageous in large scale electronic
structure calculations.  Examples of generic functions are plane waves
and and gaussians, which are often used in electronic structure
calculations.\refto{PlaneWave} Evidently, the ease of computing
Hamiltonian matrix elements with such analytical bases outweighs their
disadvantages, namely, non-locality in the case of plane waves or
non-orthogonality in the case of gaussians.  Generic basis functions
would appear to be especially desirable for vector or parallel
computations.  In contrast, an important disadvantage of energy band WF
is that they are not generic; their form depends on a given atomic
environment and hence, they must be recomputed in the course of
self-consistent calculations.

When the size of the basis becomes large, however, it becomes
computationally advantageous to trade simplicity of matrix element
construction in favor of a smaller, if more complicated basis.
Sparseness of the Hamiltonian matrix is also a significant advantage.
Yet another consideration is completeness.  In the case of gaussians, it
is not always clear whether the largest bases that can be implemented
are sufficiently complete to cover the Hilbert space of interest.  On
the other hand, gaussian basis functions may be centered on arbitrarily
positioned atoms, allow for spherical symmetry, and can represent well
the cusps in atomic wavefunctions due to the divergence in the Coulomb
potential.  Mixed basis sets, \eg a combination of gaussians and plane
waves,\refto{Mixed} have also been used.  In general, such
non-orthogonal bases can become overcomplete, as the finite basis is
almost linearly dependent, and care must be taken to avoid numerical
difficulties.

These considerations suggest that desirable basis functions for large
scale electronic structure calculations should be orthogonal, complete,
generic, local and possess useful analytic properties.  To minimize
basis set size, their locality should apply both to position and
momentum space.  A set of functions which has all of these desired
properties are the generalized Wannier functions\refto{Wilson}
$w_{l,n}(x)$ introduced by one of us.  Here $l$ and $n$ are momentum and
site labels, respectively.  In this paper we discuss the potential
applicability of these Wannier functions, hereafter referred to as
``phase space Wannier functions'' (PWF), to electronic structure
calculations.  Like plane waves, the PWF form a complete set and have
simple matrix elements with respect to the kinetic energy operator
$p^2=-\nabla^2$, thereby simplifying the calculation of Hamiltonian
matrix elements.  Like gaussians, they are well localized in position
space, and matrix elements with respect to simple functions, \eg
$1,~x,~x^2,~\dots$ may be stored.  Moreover, they are shown below to be
localized exponentially in both position and momentum space.  This leads
to a sparse Hamiltonian matrix and an overall basis set size which can
be significantly smaller than that with plane waves.  These properties
suggest that the PWF functions have significant advantages in large
scale electronic structure calculations, and we argue that they are a
potential replacement or supplement to plane wave and gaussian basis
sets.  These considerations have been checked by carrying out a detailed
one-dimensional, model calculation with the PWF with convergence checks
and comparisons with other approaches.  However, a full scale test of
the basis, \eg with pseudopotentials and self-consistency remains to be
done.

The paper is organized as follows: In section II the basic properties of
the functions $w_{l,n}(x)$ are given, together with an algorithm for
their construction and a demonstration of their exponential
localization.  Some generalizations are also discussed.  In section III
the utility of these Wannier functions is illustrated with a simple
one-dimensional model.  Criteria for choosing the basis set size and
lattice constant are determined, and convergence of the ground state
energy with respect to basis size is evaluated.  Section IV contains a
summary and conclusions.

\endpage

\head{II. Properties} \taghead{2.}

In this section we summarize the properties of the phase space Wannier
functions (PWF) introduced in \ref{Wilson}.  In addition some
generalizations are introduced and their localization properties in
position and momentum space are established.

\subhead{A. Definitions}

By construction the PWF $w_{l,n}(x)$ are uniform translations of real
functions $w_l(x)$, localized about the sites of a regular lattice
$x=na$ in position space
        $$
w_{l,n}(x) = w_l(x-2\pi n). \tag WanX
        $$
Here $a$ is an arbitrary lattice constant, which for convenience is set
to $2\pi$, $l$ is a non-negative integer momentum index, as discussed
below, and $n$ is an integer site index; the functions $w_l(x)$ are of
even(odd) parity for $l$ even(odd).  For other choices of $a$,
normalized PWF are given by scaling the above functions:
        $$
w_{l,n}(x;a)=(2\pi/a)^{1/2}\,w_{l,n}(2\pi x/a). \tag WanA
        $$
The functions $w_{l,n}$ are orthogonal and normalized such that their
inner product is a delta function:
        $$
\langle w_{l,n}|w_{l',n'}\rangle = \delta_{ll'}\delta_{nn'}.  \tag Ortho
        $$
The set of $w_{l,n}(x)$ constructed in \ref{Wilson} are illustrated in
Fig.~1(a) for $0 \leq l \leq 2$.

In momentum space, Eq.\ \(WanX) becomes
        $$
\tw w_{l,n}(p) = e^{-ip2\pi n} \tw w_l(p). \tag WanP
        $$
where $\tw w_{l,n}(p)$ is the Fourier transform of $w_{l,n}(x)$.  The
function $\tw w_l(p)$ is peaked bimodally in momentum space, as
illustrated in Fig.~1(b).  More precisely, $\tw w_l(p)$ is a real(pure
imaginary) function of even(odd) parity for even(odd) $l$, and is peaked
near $p=\pm \oq (2l+1)$ (except for $l=0$ which is peaked only at the
origin).  Each function $\tw w_l(p)$ roughly covers the disjoint
intervals $[\oh l,\oh (l+1)]$ and $[-\oh (l+1),-\oh l]$ in $p$, decaying
exponentially rapidly into neighboring intervals.  Thus one may imagine
that phase space $[-\infty < x < \infty, -\infty < p < \infty]$ may be
spanned in terms of ``fuzzy'' rectangular blocks each represented by a
function $w_{l,n}$ either in position or momentum space (Fig.~2),
whence our description ``phase space Wannier functions.'' This
localization property permits the basis to be tailored at a given site
$n$ simply by increasing the number of momentum states $l_{max}(n)$ that
are used without altering the form of the functions, changing the
lattice constant $a$, or warping space.

Several additional properties are needed to define the PWF uniquely.
For example the functions of \ref{Wilson} are defined to have simple
matrix elements with respect to the kinetic energy operator
$p^2=-d^2/dx^2$,
        $$
\langle w_{l,n}|p^2|w_{l',n'}\rangle =
\cases{%
  \alpha_l \  (\delta_{n,n'-1}-\delta_{n,n'+1}),    & $l=l'+1 $;\cr
  \beta_{l,n-n'},                                   & $l=l'   $;\cr
  \alpha_{l'} \  (\delta_{n,n'+1}-\delta_{n,n'-1}), & $l=l'-1 $;\cr
} \tag P2
        $$
where\refto{Wilson} $\alpha_l=l/(4\pi)$ and $\beta_{l,n}$ decays
exponentially rapidly with $\vert n\vert$.  The $\beta_{l,n}$ are
computed numerically in momentum space using Eq.\ \(P2).  Choices of
dispersion other than $p^2$ will be discussed below.

The three-dimensional generalizations of these functions are, as in the
case of plane waves, Cartesian products, which are labelled by vector
indices, $\vec l$ and $\vec n$,
        $$
w_{\vec l\vec n}(\vec r) =
   w_{n_x,l_x}(x)w_{n_y,l_y}(y)w_{n_z,l_z}(z). \tag Cartesian
        $$
The product form of Eq.\ \(Cartesian) can considerably simplify the
computation of matrix elements; for example, matrix elements of
separable operators such as $\exp(-\lambda r^2)$ factor into one
dimensional products.  Thus matrix elements of the Coulomb potential can
be computed as a single one-dimensional integral of a product of three
one-dimensional matrix elements by use of the identity
$(1/r)=(2\pi^{-1/2}) \int_0^{\infty} d\kappa \exp(-\kappa^2r^2)$.

The functions $w_{l,n}(x)$ have a number of other useful analytical
properties.  For example the coefficients $c_{l,n}$ in the Wannier
function expansion,
        $$
f(x)=\sum_{l,n} c_{l,n}w_{l,n}(x), \tag Expansion
        $$
of a function $f(x)$ with a Taylor expansion about $x=na$ are related to
the Taylor series coefficients of $f(x)$ at $na$ as follows:
        $$
\eqalignno{
   c_{0,n} = & a^\oh\, (f(na) + \oh f''(an)({a\over{2\pi}})^2 +
                           O(f^{(iv)}(na)a^4))               \cr
   c_{1,n} = & a^\oh\, (-f'(an)({a\over{2\pi}}) +
                           O(f'''(an)a^3)         & (Taylor) \cr
   c_{2,n} = & a^\oh\, (2f''(an)({a\over{2\pi}})^2 +
                           O(f^{(iv)}a^4)).                  \cr }
        $$
Equation \(Taylor) follows from the relation between the moments
of $w_{l,n}(x)$ and the derivatives of $\tw w_{l,n}(p)$ at $p=0$, as
discussed in Appendix A.

As a consequence, the expansion of the constant function requires only
the $l = 0$ PWF (at all sites); the expansion of a linear function
requires only the $l = 0$, and $1$ functions (at all sites); and so on.
This result holds {\it locally\/} as well: in the neighborhood of a site
$x = na$ the first two PWF's ($l = 0, 1$) adequately represent an
arbitrary function as long as it does not vary rapidly on the scale of
$a$.  For sufficiently narrow cells, these functions therefore serve as
a complete set of ``shape functions,'' as in the finite element method
of \ref{finite}.  While the shape functions of the finite element method
can be more compact, their finite range leads to poor convergence in
momentum space; also, these shape functions are polynomial complete only
to some low order.

\subhead{B. Algorithm for Construction of $w_{l,n}$}

An algorithm together with a {\it Gibbs description\/} of a computer
code for constructing PWF $w_{l,n}$ with the above properties has been
given in \ref{Wilson}.  In this Section we present an alternative
procedure which has the following advantages: The Wannier functions
generated can be shown to form a complete and orthogonal basis, and the
new procedure can be generalized to make use of dispersions other than
$p^2$.  Lastly, a recursion relation for the PWF is discussed.

To prove completeness, and orthogonality of the PWF basis, we show that
the PWF's are the eigenfunctions of a Hermitian matrix $H$ representing
the operator $p^2 = - d^2/dx^2$.  The diagonal matrix elements of $H$,
which are determined from the $\beta_{l,n}$, are not known beforehand,
but are computed by an iterative method.

The site-translation property of Wannier functions allows one to work in
a reciprocal space description in which $p = m + q$ with $m$ an integer,
and $q$ in the first Brillouin zone.  In this picture the orthogonality
relation, Eq.\ \(Ortho) becomes
        $$
\sum_n \langle w_{l,0}|w_{l',n}\rangle e^{2\pi i q n} =
\sum_{m = -\infty}^{+\infty}
        \tw w^*_l(m+q) \tw w_{l'}(m+q) = \delta_{ll'}.  \tag OrthoRecSpace
        $$
Similarly, defining $H_{l,l'}(q) = \Sigma_n \langle w_{l,0} | p^2 |
w_{l',n} \rangle e^{2\pi i q n}$, the $p^2$ matrix elements, Eq.\ \(P2),
become
        $$
H_{l,l'}(q) \equiv \sum_{m = -\infty}^{+\infty}
\tw w^*_l(m+q) \tw w_{l'}(m+q)\, (m+q)^2 = \cases{
    2i\alpha_l\,\sin 2\pi q,                              & $l=l'+1$;  \cr
      \beta_l(q), & $l=l'$;    \cr
   -2i\alpha_{l'}\,\sin 2\pi q,                           & $l=l'-1$;  \cr
    }  \tag P2RecSpace
        $$
with $\alpha_l = l/(4\pi)$, and $\beta_l(q)\equiv\Sigma_n \beta_{l,n}
e^{2\pi i nq}$.

For each $q$, the tridiagonal matrix $H_{l,l'}(q)$ defines an eigenvalue
problem for which the eigenvalues are known explicitly, and the
eigenfunctions are the desired PWF.
        $$
\sum_{l' = 0}^{\infty} H_{l,l'}(q)
\tw w_{l'}(m+q) = (m+q)^2\tw w_l(m+q).\tag Eigen
        $$
The diagonal elements of $H_{l,l'}(q)$ are computed numerically as follows:
First, the range of $l$ is made finite by limiting it to $\{0, \cdots,
L\}$ with $L$ even.  The set of eigenvalues is also truncated to
$\{(-\oh L +q)^2, (-\oh L + 1 + q)^2, \cdots, (\oh L + q)^2\}$ to
reflect the truncated momentum space.  The truncation introduces
significant error in quantities near the upper end of momentum space;
for example, the values of $\beta_L(q)$, $\tw w_L(m+q)$, and $\tw
w_0(\oh L+q)$ computed in this approximation differ significantly from
their true values.  On the other hand, the effects of truncation diminish
rapidly as one moves away from the upper end, and accurate $\beta_l(q)$
can be computed for $l$ small compared to $L$.

We evaluate the characteristic equation for each of the $L+1$ eigenvalues
to obtain a set of equations determining the $L+1$ unknowns $\beta_l(q)$,
        $$
\hbox{det}\vert H_{l,l'}(q) - (m+q)^2 \delta_{l,l'}
 \vert = 0,\quad\hbox{for\ } m
               = \{-\oh L, \cdots, \oh L\}. \tag Character
        $$
These equations may be solved by a fixed-point iterative scheme as each
$\beta_l(q)$ may be readily expressed in terms of the others because the
determinant is linear in each $\beta_l(q).$\refto{Sullivan}  The expected
bimodality of the $\tw w_l(p)$ suggests that the starting values for the
$\beta_l(q)$ be chosen to be near $(\oh l + q)^2$ for $l$ even, and
$(-\oh (l+1) + q)^2$ for $l$ odd.  Once the $\beta_l(q)$ are known, a
quick diagonalization yields the $\tw w(m+q)$.

Note that the method of \Ref{Wilson} was to solve for the $\tw w_l(p)$
without computing the $\beta_l(q)$ first.  This is more efficient and is
the preferred method for generating the PWF basis; however, the present
prescription can also be generalized to dispersions other than $p^2$ as
discussed in Section II C, below.

Recursion relations between the functions $\tw w_l(p)$ are also 
useful.  From the matrix elements in Eq.\ \(P2) one has
        $$
\alpha_{l+1}\,\tw w_{l+1}(p) =\alpha_l\,\tw w_{l-1}(p)+
   {(p^2-\beta_l(p))\over 2i\sin2\pi p}\,\tw w_l(p),
   \qquad (l\ge0,\ \alpha_0=0),\tag Recursion
        $$
Thus, given the function $\tw w_0(p)$, the remaining functions and the
$\beta_l(p)$ can be determined in principle by recursion.  Note that the
functions $\tw w_l(p)$ have zeros at the half integers (at which the
denominator $2i\sin 2\pi p$ vanishes) with the exception of $p = \pm\oh
l$, and $p = \pm\oh(l+1)$ (at which the numerator $p^2-\beta_l(p)$ also
vanishes).  Tables of the functions $\tw w_l(p)$ are available from the
authors.

\subhead{C. Generalization to Arbitrary Dispersion}

An advantage of the construction algorithm outlined in Section\ II B is
that the generalization to other dispersions $H=\epsilon(p)$ is easily
carried out.  We consider only dispersions that do not have the
periodicity of the reciprocal lattice and are even functions of $p$.
Under these restrictions the generalized WF can be chosen to have
definite parity and cover the phase space in the same manner as before.

Matrix elements of the dispersion are again required to have a
tridiagonal form: Eqs.\ \(P2RecSpace) and \(Eigen) are replaced by
        $$
H_{l,l'}(q) \equiv \sum_{m = -\infty}^{+\infty}
        \tw w^*_l(m+q) \tw w_{l'}(m+q)\, \epsilon(m+q) = \cases{
    2i\alpha_l\,\sin 2\pi q,       & $l=l'+1$;  \cr
      \beta_l(q),                  & $l=l'$;    \cr
   -2i\alpha_{l'}\,\sin 2\pi q,    & $l=l'-1$;  \cr
    }  \tag NewP2RecSpace
        $$
with $ \beta_l(q)\equiv\Sigma_n \beta_{l,n} e^{2\pi i nq}$ and
        $$
\sum_{l' = 0}^{\infty} H_{l,l'} w_{l'}(m+q) = \epsilon(m+q) w_l(m+q).
             \tag NewEigen
        $$
By an analysis similar to that of \ref{Wilson} we find\refto{Note} that
the coefficients $\alpha_l$ are given by $\epsilon'(\oh l)/4\pi$, where
$\epsilon'(p)=d\epsilon(p) / dp$.  The remaining steps of the
construction procedure are carried out as in the previous
section.  The iteration is started with the $\beta_l(q)$ chosen to be
near $\epsilon(\oh l + q)$ for $l$ even, and $\epsilon(-\oh(l+1)+q)$
for $l$ odd.  The effect of dispersion on the PWF is illustrated in
Fig.~3, where functions for the dispersions $\epsilon(p)=p^2$ and
$\epsilon(p)=p^4$ are compared.

\subhead{D. Exponential Localization}

The Wannier functions $w_{l,n}$ of \ref{Wilson} appear to be localized
exponentially in position space and their fourier transforms in momentum
space.  This is surprising in view of a theorem proved independently by
Balian and by Low,\refto{BalianLow} which shows that it is impossible to
construct Wannier functions which are exponentially localized about the
sites of a regular lattice in phase space with cell area $2\pi$.  In
this section we discuss how this theorem can be evaded to construct
functions which do possess exponential localization.

The non-existence theorem of \ref{BalianLow} applies specifically to
functions which are uniform translations in phase space of a singly
peaked function $f(p)$, as in the functions of von
Neumann,\refto{VonNeumann}
        $$
g_{l,n}(p)= f(p-l) e^{-i2\pi pn}, \tag CoherentState
        $$
for integers $l,n$.  As a consequence expansions in ``coherent states,''
where $f(p)$ is a gaussian, converge very slowly (\ie algebraically), so
their usefulness in  quantitative calculations is limited.  By
contrast, the PWF are bimodal in phase space.  They are composed of
functions peaked about $p=\pm \oq (2l+1)$, as in
        $$
g_{l,n}(p)= \left[{f(p-\oq (2l+1))+(-1)^l f(-p-\oq (2l+1))}\right]
   e^{-i2\pi pn}, \tag BiModal
        $$
for integers $l\ge0$ and $n$.  This bimodal form is sufficient to render
the theorem in \Ref{BalianLow} inapplicable.  Strictly speaking, the
bimodal form of $g_{l,n}(p)$ implies that the rms width of such
functions in momentum space is $O(l)$ for large $l$; but the two peaks
$f(p)$ in Eq.\ \(BiModal) are each well localized about $|p|=\oq
(2l+1)$.

To see how bimodality of $g_{l,n}$ leads to exponential localization, we
consider the problem of generating a complete set of orthonormal,
Wannier functions $w_{l,n}$ starting from {\it trial functions\/} of the
form of Eq.\ \(BiModal), where $f(p)$ is a normalizable, even function
assumed to be well localized about the origin.  The functions $w_{l,n}$
are constructed \refto{Kohn} by a symmetric orthogonalization of the
functions $g_{l,n}$, using the inverse square root of the overlap matrix
$S_{ln,l'n'}=\langle g_{l,n}\vert g_{l',n'}\rangle:$
        $$
w_{l,n}=\sum_{l',n'} (S^{-1/2})_{ln,l'n'}\, g_{l',n'}. \tag ReOrth
        $$
The matrix $S^{-1/2}$ can be calculated by going to a diagonal
representation in terms of the solutions to the eigenvalue problem,
        $$
\sum_{l',n'}\, S_{ln,l'n'}\, c_{l',n'} = \lambda c_{l,n}. \tag SEigen
        $$
This construction is carried out in Appendix B where it is shown that
$(S^{-1/2})_{ln,l'n'}$ decays exponentially both in $|n-n'|$ and in
$|l-l'|$ for large $l$.  This result implies exponential
localization\refto{RehrKohn} with the same decay constants for the
Wannier functions, $w_{l,n}$ defined by Eq.\ \(ReOrth), for sufficiently
localized trial functions $f(p)$ (\eg gaussians).  Moreover, the
translational invariance $(S^{-1/2})_{ln,l'n'}$ in $l$ for large $l$
implies that the Wannier functions in momentum space will be nearly
identical for large $l$, a form which is already evident from Fig.~1(b).

\endpage

\head{III. Application to Quantum Mechanics in One-dimension} \taghead{3.}

\subhead{A. Schrodinger Equation in the PWF Basis}

As an illustration of their properties and the application of the PWF we
use them as a basis for solving the one-dimensional Schrodinger equation
        $$
H\psi=
   \left[{
 -{d^2 \over dx^2} + V(x)}\right]\psi =
   E \psi. \tag Schroedinger
$$
We take $V(x)$ to be an attractive single-well potential of the form
        $$
V(x) = -\lambda\  \hbox{sech}^2(x), \tag Potential
        $$
with $\lambda = 35/4$ (Fig.~4).  For this choice, the bound state solutions
$\psi(x)$ are known analytically, so that precise comparisons are
possible: there are three bound states of which the lowest is at $E_0 =
-25/4$, and $\psi_0=(8/3\pi)^{1/2}\hbox{sech}^{5/2}(x)$.

In the PWF basis, $\psi(x)=\Sigma_{ln}\, c_{ln}\, w_{ln}(x)$, and the
Schrodinger equation \(Schroedinger) is equivalent to the matrix
eigenvalue problem,
        $$
\sum_{l'n'} H_{ln,l'n'}\, c_{l'n'} = E\, c_{ln}. \tag MatrixSchroed
        $$
It is convenient to compute the kinetic energy part of $H$ in momentum
space using Eq.\ \(P2) and the potential energy part in position space,
so that
        $$
H_{ln,l'n'}= \oh\langle w_{ln}|p^2|w_{l'n'}\rangle +
                \langle  w_{ln}|V(x)|w_{l'n'}\rangle. \tag Ham
        $$
The sum over $n'$ in Eq.\ \(MatrixSchroed) is restricted to box of
length $L$ so that $-\oh L \leq x=na \leq \oh L$, for a given choice of
$L$ and lattice constant $a$, while that over $l'$ is restricted to $l'
\leq l_{max}(n')$.  The exact coefficients $c_{l,n}$ can be obtained
conveniently in momentum space,
        $$
c_{l,n} = \int dp\, w_{l,n}^*(p) \psi_0(p), \tag Coeff
        $$
where $\psi_0(p)= (8/3\pi)(2/3\pi)^{1/2}
|\Gamma({\scriptstyle{5\over4}}+\oh ip)|$.  These coefficients are given
in Table 1 for lattice constants $a=0.5, 1.5$, and 3.5, and a fixed
overall length $L=10.5$, a value chosen to be a few times the width of
the potential well $V(x)$.  Choosing, for example, 0.01 as a cutoff for
the phase-space cell occupations $|c_{l,n}|^2$, the representation of
the ground state wave function requires 5 PWF, \ie
($l=0;n=0,\pm1,\pm2$), for lattice constant $a=0.5$; 3 PWF, \ie
($l=0;n=0,\pm1$), for $a=1.5$; and just 2 PWF, ($l=0,2,n=0$), for
$a$=3.5.  For comparison, a plane wave basis to the same $1\%$ accuracy
requires 9 plane waves, $k=2\pi m/L$ with $-4 \le m \le 4$ (see Table
2).  With the cutoff set at 0.003, the corresponding numbers are 7
Wannier functions for $a$ = 0.5, 5 for $a$ = 1.5, 6 for $a$ = 3.5, and
11 plane waves.  Note that the example we have chosen is centered at
$x=0$, so some coefficients vanish automatically; similarly, if
symmetrized plane waves were used, only 5 (or 6 for a cutoff of 0.003)
even functions would be needed.  However, in practice one cannot assume
symmetry.  Without symmetry, the $n=0,l=1$ amplitudes need not vanish,
increasing the counts for the Wannier functions by one for all three
values of $a$; similarly one needs all plane waves.  Note also that the
number of plane waves needed depends on the length of the system $L$,
while the number of Wannier functions is determined locally and is
independent of $L$.  In the present example, the number of plane waves
can be reduced by cutting down $L$, but this restricts the location of
the orbital.

\subhead{B. Phase Space Criteria for Basis Set Size}
We now discuss the dependence of the coefficients $c_{l,n}$ on $l$ and
$n$ from phase-space considerations.  To the extent that a
semi-classical description is valid, these coefficients should be small
for all phase space cells which do not overlap the classical trajectory.
This indeed appears to be the case and suggests a criterion for adopting
a given lattice constant and basis set size.  One choice of the lattice
constant $a$ is that value for which the ``coverage'' $\sum_{occ}
|c_{l,n}|^2$ is optimal, \ie the value for which the minimum number of
phase space cells overlap the classical region of occupied states.  From
Table 1 the coverages at a cutoff of 0.01 in $|c_{l,n}|^2$ are 0.978,
0.985 and 0.975 for lattice constants $a$ = 0.5, 1.5, and 3.5,
respectively.  For a cutoff of 0.003, the corresponding figures are
0.990, 0.999 and 0.997.  For both cutoff choices, the value $a=1.5$
maximizes the coverage with a minimum number of functions and is
therefore optimal for the values being considered.  Inspection of Fig.~5
indicates that this lattice constant corresponds roughly to the size of
the phase space orbit.  Were the lattice constant chosen to differ
greatly from this value, the phase space cells would not ``tile" the
orbit efficiently, and more PWF's would be needed to achieve comparable
coverage of the ground state wavefunction.

For comparison, a plane wave expansion spans phase space with elongated
cells of width $\Delta x = L$ and height $\Delta p = 2\pi/L$, and
therefore requires $2p_{max}L/2\pi$ states, for an equivalent
description.  The value of $p_{max}$ is related to the depth of a
potential well.  For example, in the case of a point charge $Ze$,
$\psi(x)\sim \exp(-Z x/a_o)$ and $p_{max}\sim Z/a_o$, where $a_o$ is the
Bohr radius.  The use of pseudopotentials, therefore, can considerably
reduce the size of basis required for a given calculation, either in a
plane-wave or a PWF basis.  Moreover, one expects from general phase
space volume considerations (\ie from the uncertainty principle) that
the number of Wannier functions needed will be of order 2$\int dx\,
p(x)/2\pi\hbar$ which is just the area of phase space within the
classical orbit at energy $E$, in units of $2\pi\hbar.$ The Wannier
functions are therefore expected to be more efficient than the plane
waves by a factor roughly equal to the ratio of these phase-space areas
(phase-space volumes in three dimensions).  It is also of interest to
examine the convergence of the ``coverage'' {\it vs.\/} total basis
size.  For smooth potentials, the convergence of the coverage with
respect to $l$ at a given $n$ appears to be exponentially rapid, which
is what one expects based on WKB arguments.  For the one-dimensional
example discussed above, it is observed that the number of plane waves
required to give an accuracy of $1\%$ (or $0.3\%$) in the coverage is,
ignoring symmetry, about 9/6 (or 11/6) times that for the Wannier
functions; in three dimensions the corresponding factor would be cubed,
\ie with the plane wave basis larger by a factor of 3 to 6, depending on
the choice of cutoff.  Thus because it can be adjusted locally, the PWF
basis can be significantly smaller than a plane wave basis.

\subhead{C. Convergence of Ground State Energy}

Finally, we have computed the coefficients $c_{l,n}$ and the eigenvalue
$E_0$ for the lowest state by straightforward diagonalization of the
Hamiltonian matrix in Eq.\ \(MatrixSchroed) for several different basis
set sizes.  The results are shown in Table 3.  Note that convergence of
the ground state energy and the convergence of the coverage are
comparable since both vary quadratically with wavefunction error.  For
example, for the case $a=1.5$ with 5 basis functions the difference from
unity of the coverage, \ie $1-\Sigma |c_{l,n}|^2$, is 0.0027, and the
relative error in the ground state energy, $\Delta E_0/E_0$, is 0.0008.
For $a=3.5$ with 6 basis functions, the corresponding numbers are 0.0046
and 0.0102, respectively.  This also indicates the importance of total
coverage as a gauge of convergence of a given basis set.

\endpage

\head{IV. Conclusions} \taghead{4}

We have suggested that phase space Wannier functions may be an
attractive basis for large scale electronic structure calculations, as
they constitute an orthonormal, complete, local and generic basis set.
Their localization in position and momentum space permits the basis set
to be optimized locally, and they have many convenient properties, \eg
simple short-range matrix elements of the kinetic energy operator.  On
the other hand, we have not constructed an analytical representation of
these functions; at present, they are tabulated numerically, and
typically 100 points are needed to represent their fine structure.
Detailed properties of these functions and some generalizations have
been given, together with an algorithm for their construction and proof
of their exponential localization properties.

An illustrative application to the Schrodinger equation in one-dimension
has also been given.  The results of our one-dimensional example suggest
criteria for choosing the basis set size and lattice constant $a$ based
on semi-classical, phase space considerations.  Compared with a plane
wave basis set, the phase space Wannier functions permit a significant
reduction in basis set size, since the basis size can be adjusted locally.
Even a small reduction is significant in three dimensions, since
the number of operations required in a straightforward
matrix diagonalization varies as the cube of the basis size.

The next step would be to carry out a three-dimensional electronic
structure calculation using these functions, i.e., a calculation
analogous to a comparable plane-wave, pseudopotential calculation.
Important developments needed for self-consistent calculations are a) a
good representation of pseudopotential matrix elements, and b) an
efficient scheme for solving Poisson's equation in the PWF basis.

\goodbreak\bigskip
\leftline{\uppercase{Acknowledgments}}
\nobreak\vskip 0.25truein\nobreak

We thank G. Battle, Z. Levine, T. Olson, M. Teter, C. Umrigar and S. White
for comments and suggestions.  One of us (JJR) also thanks the
Laboratory of Atomic and Solid State Physics at Cornell University for
hospitality and support at the time this work was initiated.  This work
was supported in part by DOE -- Basic Energy Sciences, Division of
Materials Research and by NSF, DMR 87-02002.

\endpage

\head{Appendix A} \taghead{A}

\subhead{Taylor and Wannier Function Expansion Coefficients}

In this appendix we discuss the relation between the Wannier function
expansion coefficients $c_{l,n}$ of a function $f(x)$,
        $$
f(x)=\sum_{l,n} \, c_{l,n}\, w_{l,n}(x;a),\qquad
 c_{l,n} = \int dx\, w_{l,n}(x;a)\,f(x) , \tag A2
        $$
and the Taylor series expansion coefficients of $f(x)$ about the point
$x=na$ , \ie
        $$
f(x)=\sum_j \, {f^{(j)}(na)\over j!}(x-na)^j. \tag A1
        $$
Using equations \(A2), \(A1) and \(WanA), two expressions for the expansion
coefficients are readily obtained.
        $$
\eqalignno{
c_{l,n} & = (a/2\pi)^\oh \sum_j \, {f^{(j)}(na)\over j!} (a/2\pi)^j\,
           \int dx\, x^j\, w_l(x)         \cr
        & = (a/2\pi)^\oh \sum_j \, {f^{(j)}(na)\over j!} (a/2\pi)^j\,
           \mu_{l,j}              \cr
        & = a^\oh\, \sum_j \, {f^{(j)}(na)\over j!} (ia/2\pi)^j\,
           \tw w^{(j)}_l(0)       & (A4) \cr }
        $$
where $\mu_{l,j}$ is $j^{th}$ moment of $w_l(x)$, and $\tw w^{(j)}_l(p)$
is the $j^{th}$ derivative of $\tw w_l(p)$.  This last equation
follows from the equality between the moments and the derivatives of
$\tw w_l(p)$ evaluated at the origin.
        $$
\mu_{l,j}=(2\pi)^\oh\,i^j \tw w_l^{(j)}(0). \tag A5
        $$

Now: the sum in Eq.\ \(A4) simplifies considerably since (a) many of the
moments vanish by symmetry arguments, and (b) many more vanish by the
structure of the $\tw w_l(p)$ at the origin.  From the recursion
relation Eq.\ \(Recursion) one can determine the location as well as the
order of the zeros of $\tw w_l(p)$.  In particular, at the origin the
following derivatives vanish,
        $$
\tw w^{(j)}_l(0) = 0 \quad \hbox{for $j < l$.} \tag A6
        $$
Further, several of the non-vanishing derivatives can be deduced from
the recursion and the orthogonality equations: $\tw w_0(0)=1$, $\tw
w_0''(0)=-1$, $\tw w_1'(0)= i$, $\tw w_2''(0)=-4$; the remaining
derivatives, if needed, can be calculated numerically.  Using Eq.\ \(A4)
and known values of the derivatives, we obtain the results given in Eq.\
\(Taylor).

\endpage

\head{Appendix B} \taghead{B}

\subhead{Exponential Decay of Phase Space Wannier Functions}

In this section we construct the inverse square root
$(S^{-1/2})_{ln,l'n'}$ of the overlap matrix $S_{ln,l'n'}=\langle
g_{l,n}|g_{l',n'}\rangle$ for a set of trial Wannier functions $g_{l,n}$
given by Eq.\ \(BiModal).  The computation is most conveniently carried out
in a diagonal representation, \ie in terms of the eigenstates
$c_{l,n}$ of the matrix $S_{ln,l'n'}$:
        $$
\sum_{l',n'} S_{ln,l'n'} c_{l',n'} = \lambda c_{l,n}. \tag B1
        $$
This calculation is carried out in several steps as follows.  First by
translational invariance in position space, we may take $c_{l,n}$ to be
of Bloch form, $c_{l,n}=c_l(k)\exp(ikn)$ so that Eq.\ \(B1) reduces to a
one-dimensional eigenvalue problem, $\Sigma_{l'}S_{l,l'}(k)c_{l'}(k)=
\lambda_kc_l(k),$ where $S_{l,l'}(k)=\Sigma_n \exp(2\pi ikn)S_{l0,l'n}.$
Second, note that by construction the form of the matrix $S_{l,l'}(k)$
depends on whether $l$ and $l'$ are even or odd.  Thus we define
$l\equiv2\mu+\sigma$, $(\sigma=0,1)$.  We find below that for large
$l,l'$ the matrix $S_{l,l'}(k)$ becomes translationally invariant in
$\mu$.  The eigenvalue problem is thus analogous to a tight-binding
problem on a semi-infinite chain with two basis states ($\sigma=0,1$)
per unit cell.  The limiting form of $S_{l,l'}(k)$ is seen to be
        $$
S_{l,l'}(k)\rightarrow S_{\sigma\sigma'}(k,\mu-\mu')=\left[{
\tilde S_{\sigma\sigma'}(k,\mu-\mu')+
(-1)^{\sigma+\sigma'}\tilde S_{\sigma\sigma'}(-k,\mu-\mu')}\right], \tag B2
        $$
where nontranslationally invariant terms which vary as $\tilde
S_{\sigma,\sigma'}(k,\mu+\mu')$ have been neglected; those terms are
negligible for sufficiently localized functions $f$.  In terms of the
functions $f(p)$ in Eq.\ \(BiModal) the quantity in Eq.\ \(B2) is
        $$
\tilde S_{\sigma,\sigma'}(k,\mu)=\sum_m f(k+m-\oq(\sigma+1))\,
f(k+m+\mu-\oq(\sigma+1)).    \tag B3
        $$
Next, the eigenvalues of Eq.\ \(B1) can be obtained from the large $l$
limit where, by translational invariance in $\mu$,
$c_{l}(k)\rightarrow\exp(\pm i2\pi q\mu)c_\sigma(k,q)$.  Then one
obtains the $2\times2$ eigenvalue problem,
        $$
S_{\sigma,\sigma'}(k,q)c_\sigma'(k,q)=\lambda_{k,q}c_\sigma(k,q), \tag B4
        $$
where
        $$
S_{\sigma,\sigma'}(k,q)=\sum_\mu S_{\sigma\sigma'}(k,\mu)
e^{iq\mu}=\tilde S_{\sigma,\sigma'}(k,q)+(-1)^{\sigma+\sigma'}
\tilde S_{\sigma,\sigma'}(-k,q). \tag B5
        $$
It is easily verified that the functions $\tilde S_{\sigma,\sigma'}$
factor as $\tilde S_{\sigma,\sigma'}(k,q)=$ $ \theta_\sigma^*(k,q)
\theta_{\sigma'}(k,q),$ where
        $$
\theta_\sigma(k,q)=\sum_m e^{i2\pi qm}f(k+m-\oq(2\sigma+1)). \tag B6
        $$
The functions in Eq.\ \(B6) are analogous to those discussed by Balian
and by Low.\refto{BalianLow} They show that $\theta_\sigma(k,q)$ must
have at least one real zero in the Brillouin zone [$0\le k\le1,0\le
q\le1$].  This property is sufficient to rule out exponential decay for
Wannier functions constructed from the singly-peaked trial functions of
Eq.\ \(CoherentState).  For the functions $\theta_\sigma(k,q)$ one can also
establish the identity $\theta_1(k,q)=\exp(i2\pi q)\theta_0^*(-k,q)$, from
which it follows that the $2\times2$ matrix $S_{\sigma,\sigma'}(k,q)$ is
diagonal, with diagonal elements given by
$|\theta_0(k,q)|^2+|\theta_1(k,q)|^2$.  Hence, barring the unlikely
possibility of common zeroes both in $\theta_0$ and $\theta_1$, the
eigenvalues are positive definite for real $k,q$.  For gaussian trial
functions $f(p)$, for example, there is only a single zero of
$\theta_0(k,q)$ at $q={1\over2}$ and $k={3\over4}$.

Finally, the inverse square root matrix is obtained by quadrature, and
we obtain
        $$
(S^{-1/2})_{ln,l'n'}=\int\!\!\!\int_{BZ} dk\, dq \,\sum_{\alpha=0,1}
{c^\alpha_l(k,q) c^\alpha_{l'}(k,q) \over
[|\theta_0(k,q)|^2+|\theta_1(k,q)|^2]^{1/2}}
e^{i2\pi k(n-n')},   \tag B7
        $$
where $c_l(k,q)=A_l(k,q) \cos(q\mu+\delta_l(k,q))$ are the exact
eigenstates, $\delta_l(k,q)$ being a phase shift which is $l$ dependent
for small $l$.  Since the integrand has no zero for real $k,q$ in the
Brillouin zone, $(S^{-1/2})_{ln,l'n'}$ will decay exponentially in
$|n-n'|$ and in $|l-l'|$ for large $l$, at rates determined by the
nearest zeroes of $\lambda(k,q)$ in complex $k,q$ space.  Applying this
result to Eq.\ \(ReOrth), one sees that the Wannier functions $w_{l,n}$
have the same decay rates;\refto{RehrKohn} \ie there is decay constant
$\bar h_n$ such that $\exp(hn)w_{l,n}\rightarrow 0, n\rightarrow \infty,
h \le \bar h_n,$ and similarly for decay with respect to $l$.  In
\Ref{Wilson} these decay constants are found empirically to be $\bar h_n
= 2.9$ and $\bar h_l = 0.46$, respectively.

\endpage

\references



\refis{Wilson} K. G. Wilson, ``Generalized Wannier Functions,"
preprint (submitted to Phys. Rev. B).

\refis{Kohn} W. Kohn, Phys. Rev. B {\bf 7}, 4388 (1973); see also,
J. des Cloizeaux, Phys. Rev. {\bf135}, A698(1964).

\refis{PlaneWave} See, for example,
{\it Computational Methods in Band Theory}, T.  H.  Dunning, Jr.  and P.
J.  Hay, in {\it Methods of Electronic Structure Theory,} ed.  H.  T.
Schaefer, (Plenum, NY 1977), p.  3.

\refis{Mixed} S. G. Louie, K. M Ho, and M. L. Cohen, Phys. Rev. B
{\bf 19}, 1774 (1979).

\refis{finite} O. C. Zienkiewicz, {\it The Finite Element Method}
(McGraw-Hill, London, 1977); see also,  S. R. White, M. P. Teter, and
J. W. Wilkins,  Phys. Rev. B {\bf 39} 5819, (1989), and references
therein.

\refis{Sullivan} We can expand the determinant in powers of one of the
$\beta_l(q)$: Setting\hb
$D[\beta_l(q)]=\hbox{det}\vert H_{l,l'}(q)-(m+q)^2 \delta_{l,l'} \vert$,
one has $0 = D[\beta_l(q)+\Delta\beta_l(q)] =
D[\beta_l(q)]+\Delta\beta_l(q) (\partial D[\beta_l(q)]/\partial
\beta_l(q)) = D[\beta_l(q)](1 + \Delta\beta_l(q) G_{l,l}(q)),$ where
$G_{l,l'}(q)$ is the inverse of $H_{l,l'}(q) - (m+q)^2 \delta_{l,l'}$.
Solving for $\Delta\beta_l(q)$, we obtain $\Delta\beta_l(q) = -1 /
G_{l,l}(q)$.  The diagonal elements of the inverse of a tridiagonal
matrix are rapidly computable.

\refis{Note} The analyticity of $\tw w_l(p)$ as $q$ approaches zero
determines $\alpha_l$: $\alpha_l = (\epsilon(q+\oh l) - \hfil\break
\epsilon(q-\oh l)) /[4 \sin 2\pi q]$.  Evaluating this expression using
l'Hospital's rule fixes $\alpha_l$ to be $\epsilon'(\oh l)/4\pi$.

\refis{BalianLow} R. Balian, C.R. Acad. Sci. Paris {\bf292},
1375 (1981); F. E. Low, J. Math. Phys. {\bf 28}, 1089 (1987); and
F. E. Low in {\it A Passion for Physics, Essays in Honor of Geoffrey Chew},
C. deTar, ed.  (World Scientific, Singapore, 1985); for an alternative
proof, see G. Battle, Lett. Math. Phys. {\bf15}, 175 (1988).

\refis{VonNeumann} J. von Neumann, Z. Phys. {\bf57}, 30 (1929); H.
Bacry, A. Grossman and J. Zak, Phys. Rev. B {\bf 12}, 1118 (1975).

\refis{RehrKohn} J. J. Rehr and W. Kohn, Phys. Rev. B {\bf 10}, 448 (1974).

\endreferences

\figurecaptions

\noindent
Fig.~1.  Phase space Wannier functions $w_{l}$ for the dispersion
relation $p^2$ with lattice constant $a=2\pi$ (a) in position space:
$l=0$, solid line; $l=1$, dotted line; and $l=2$, solid-dashed line; and
(b) in momentum space, plotted with respect to the central momentum
$(2l+1)/4$.  For odd $l$, the imaginary part is plotted.  Note that the
functions are indistinguishable for $l\geq 2$.
\medskip
\noindent
Fig.~2.  Schematic plot of phase space cells (disjoint for $l > 0$)
corresponding to peaks in $w_{l,n}$ for (a) even $l$ and (b) odd $l$.
Also shown at is the approximate shape of $w_l(p)$ in momentum
space.
\medskip
\noindent
Fig.~3.  Effect of dispersion relation $\epsilon(p)$ on the PWF; shown
are functions $w_l(p)$ in momentum space for dispersion relation $p^2$
(solid line) and $p^4$ (dashed line): $l=0$ (upper curves) and the
imaginary part of $W_l(p)$ for $l=1$ (lower curves), all with lattice
constant $a=2\pi$.
\medskip
\noindent
Fig.~4.  Potential $V(x) = -\lambda\,\hbox{sech}^2(x)$ with $\lambda=
35/4$.  Horizontal lines correspond to the three bound-state energy
levels.
\medskip
\noindent
Fig.~5.  Classical phase space orbits $p(x) = \sqrt{E_i - V(x)}$ of
bound states of the potential $V(x) = -\lambda\,\hbox{sech}^2(x)$ with
$\lambda = 35/4$.  Superposed are phase space cells for different
choices of lattice constants: a) $a=0.5$, b) $a=1.5$, and c) $a=3.5$.
Tables of coefficients for these choices of $a$ are given in Table 1.

\endfigurecaptions
\endpage

\centerline{TABLE CAPTIONS}

\medskip
\noindent
Table 1.  Contribution to the ground state wave function $|c_{l,n}|^2$
from the phase space cell ${l,n}$, for different choices of lattice
constant: a) $a=0.5$; b) $a=1.5$; and c) $a=3.5$.
\medskip
\noindent
Table 2.  Contribution to the ground state wave function $|c_k|^2$ for a
plane wave expansion in the interval $-L/2\le x\le L/2$, with $L=10.5$
and $k=2\pi m/L$, for $m=0,\pm1,\pm2,\cdots\pm6$.
\medskip
\noindent
Table 3.  Ground state energy $E_0$ for various basis set sizes
$\{l,n\}$.

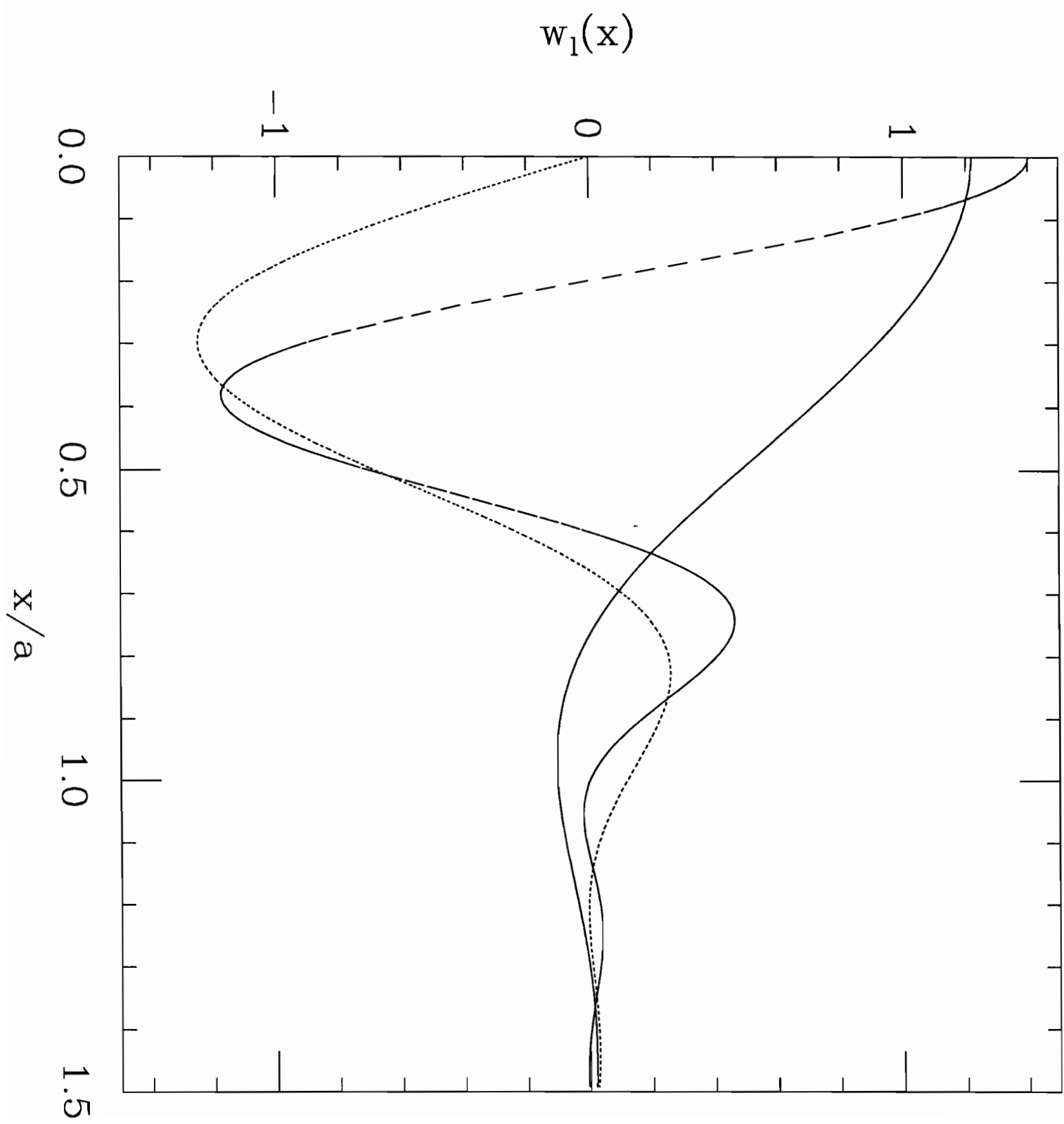
\endpage


\thicksize=0.8pt        
\def\t#1#2{$#1\times 10^{-#2}$}


\vbox{
\centerline{Table 1.}
\medskip
\begintable
        | &\multispan{4}\tstrut\hfil $\vert c_{l,n}\vert ^2$\hfil\nr
\multispan{1}|\multispan{5}\leaders\hrule height .8pt depth 0pt \hfil\nr
        | $l \setminus n\ $ %
                     | 0       | 1       | 2       | 3      \cr
\multispan{6}\cr
        | 0          | 0.41400 | 0.23226 | 0.04984 | \t{6}{3}\nr
\multispan{1}|\multispan{5}\leaders\hrule height .8pt depth 0pt \hfil\nr
a = 0.5 | 1          | 0.00000 | \t{3}{3}| \t{1}{3}| \t{2}{4}\nr
\multispan{1}|\multispan{5}\leaders\hrule height .8pt depth 0pt \hfil\nr
        | 2          | \t{2}{4}| \t{3}{5}| \t{2}{5}| \t{2}{5}\cr
\multispan{6}\cr
        | 0          | 0.92235 | 0.03112 | \t{1}{6}| \t{5}{7}\nr
\multispan{1}|\multispan{5}\leaders\hrule height .8pt depth 0pt \hfil\nr
a = 1.5 | 1          | 0.00000 | 0.00731 | \t{4}{5}| \t{1}{6}\nr
\multispan{1}|\multispan{5}\leaders\hrule height .8pt depth 0pt \hfil\nr
        | 2          | \t{1}{4}| \t{1}{6}| \t{1}{6}| \t{1}{6}\cr
\multispan{6}\cr
        | 0          | 0.84512 | \t{7}{4}| \t{1}{5}|\t{3}{8}\nr
\multispan{1}|\multispan{5}\leaders\hrule height .8pt depth 0pt \hfil\nr
a = 3.5 | 1          | 0.00000 | \t{5}{3}| \t{1}{5}|\t{2}{8}\nr
\multispan{1}|\multispan{5}\leaders\hrule height .8pt depth 0pt \hfil\nr
        | 2          | 0.12984 | \t{6}{3}| \t{3}{5}|\t{7}{8}
\endtable
}

\vfill

\noncenteredtables
\line{


\vbox{
\hbox{Table 2.}
\medskip
\begintable
$k$ | $\vert c_k\vert ^2$ \cr
 0.0000 | 0.24702 \cr
 0.5984 | 0.20007 \cr
 1.1967 | 0.11042 \cr
 1.7952 | 0.04524 \cr
 2.3936 | 0.01500 \cr
 2.9920 | 0.00429 \cr
 3.5904 | 0.00111
\endtable
}
\hfil
\vbox{
\hbox{Table 3.}
\medskip
\begintable
         |  N | $(l;n)$ | $E_0\{l,n\}$ \cr
\multispan{3}\cr
 a = 0.5 | 5 | $(0; 0, \pm 1, \pm 2)$                  | -4.8479 \nr
\multispan{1}|\multispan{3}\leaders\hrule height .8pt depth 0pt \hfil\nr
         | 9 | $(0; 0,\pm 1, \pm 2)\ (1;\pm 1, \pm 2)$ | -6.0017 \cr
\multispan{3}\cr
 a = 1.5 | 3 | $(0; 0, \pm 1)$                         | -6.0119 \nr
\multispan{1}|\multispan{3}\leaders\hrule height .8pt depth 0pt \hfil\nr
         | 5 | $(0; 0, \pm 1)\ (1; \pm 1)$             | -6.2333 \cr
\multispan{3}\cr
 a = 3.5 | 2 | $(0; 0)\ (2;0)$                         | -5.9969 \nr
\multispan{1}|\multispan{3}\leaders\hrule height .8pt depth 0pt \hfil\nr
         | 6 | $(0; 0)\ (1; \pm 1)\ (2; 0, \pm 1)$     | -6.1860
\endtable
}
}

\endit
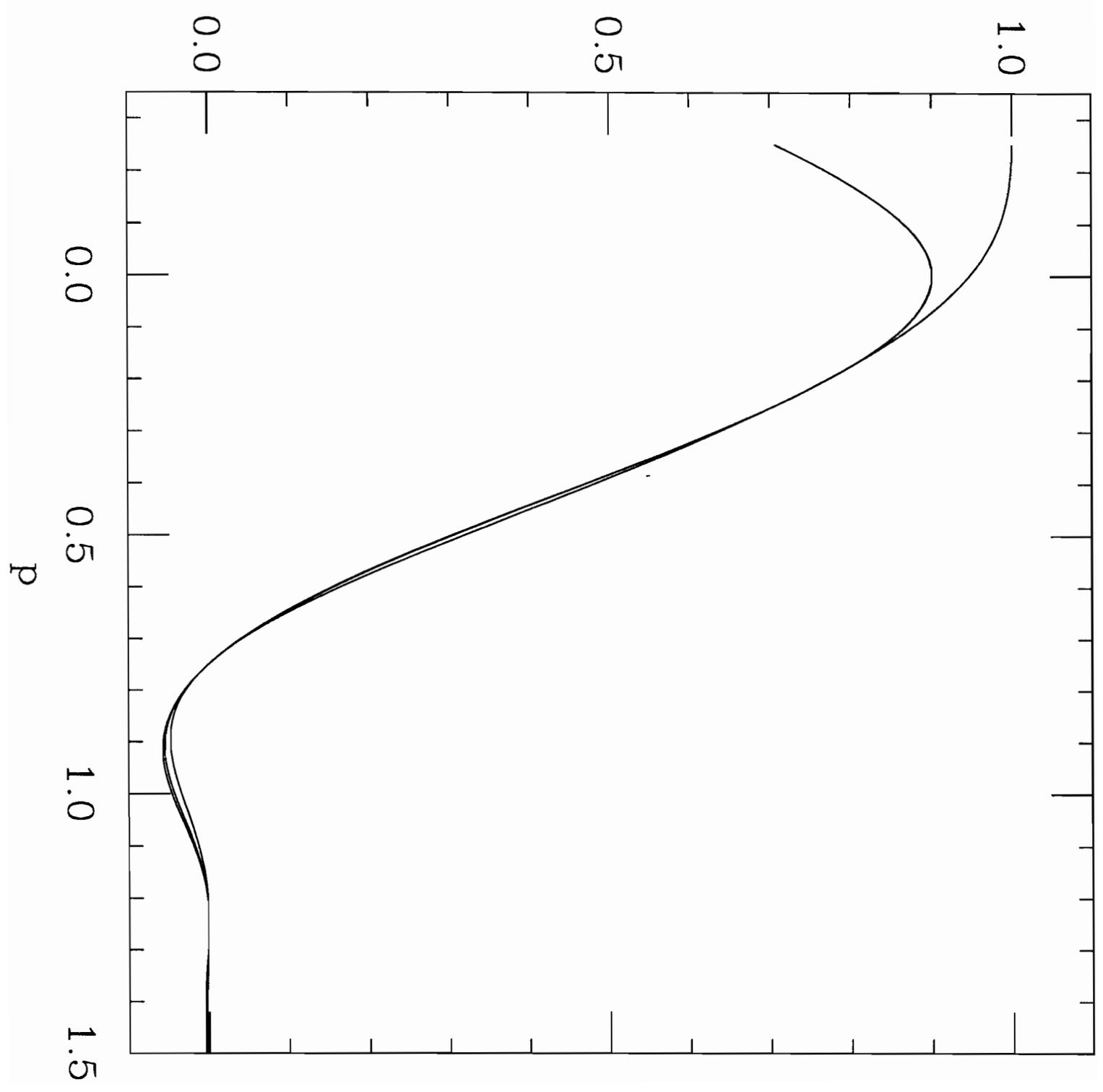
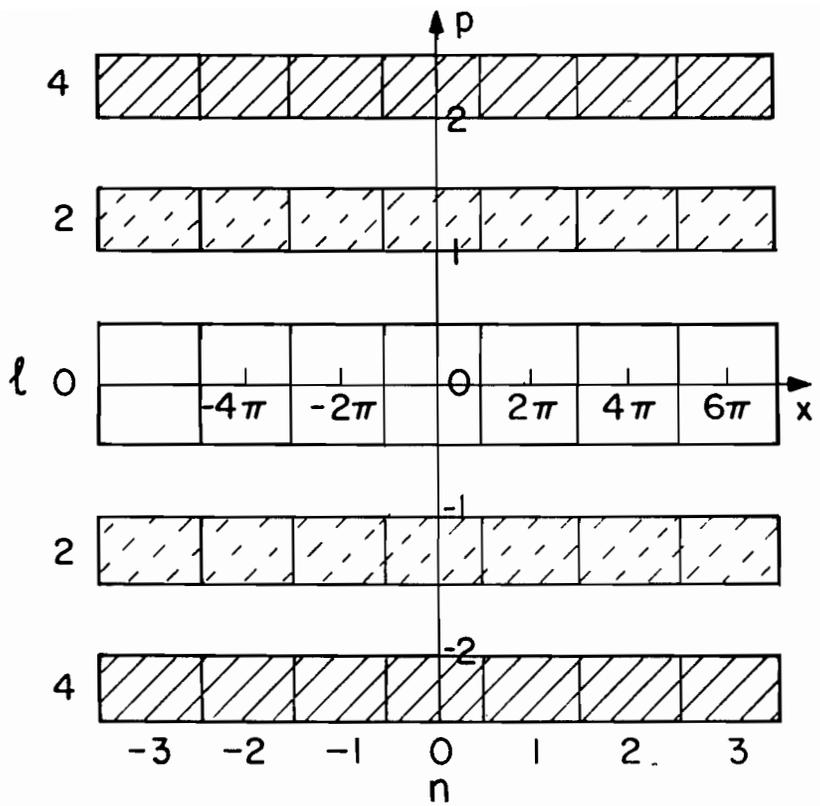
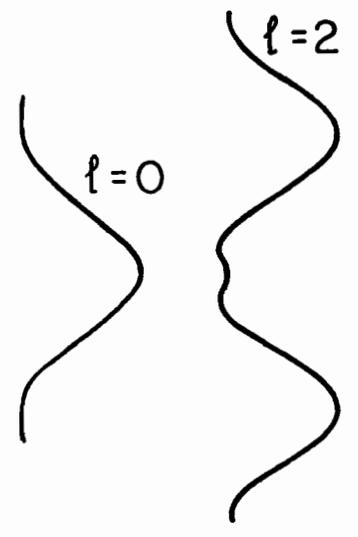
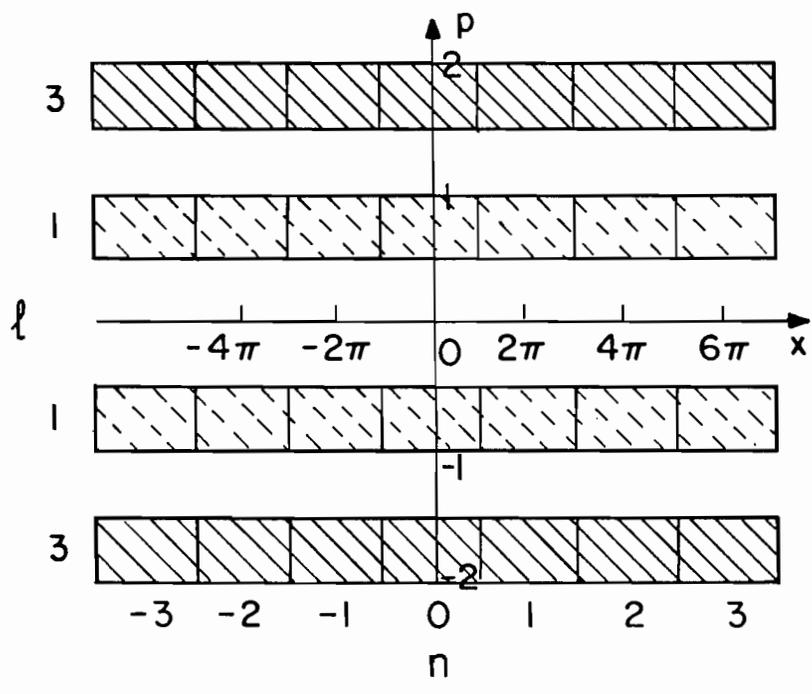
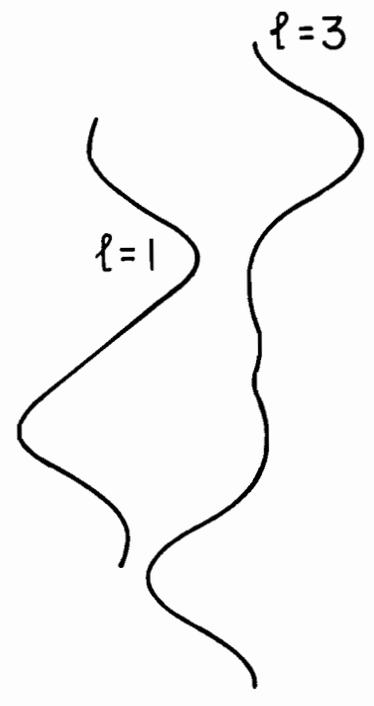
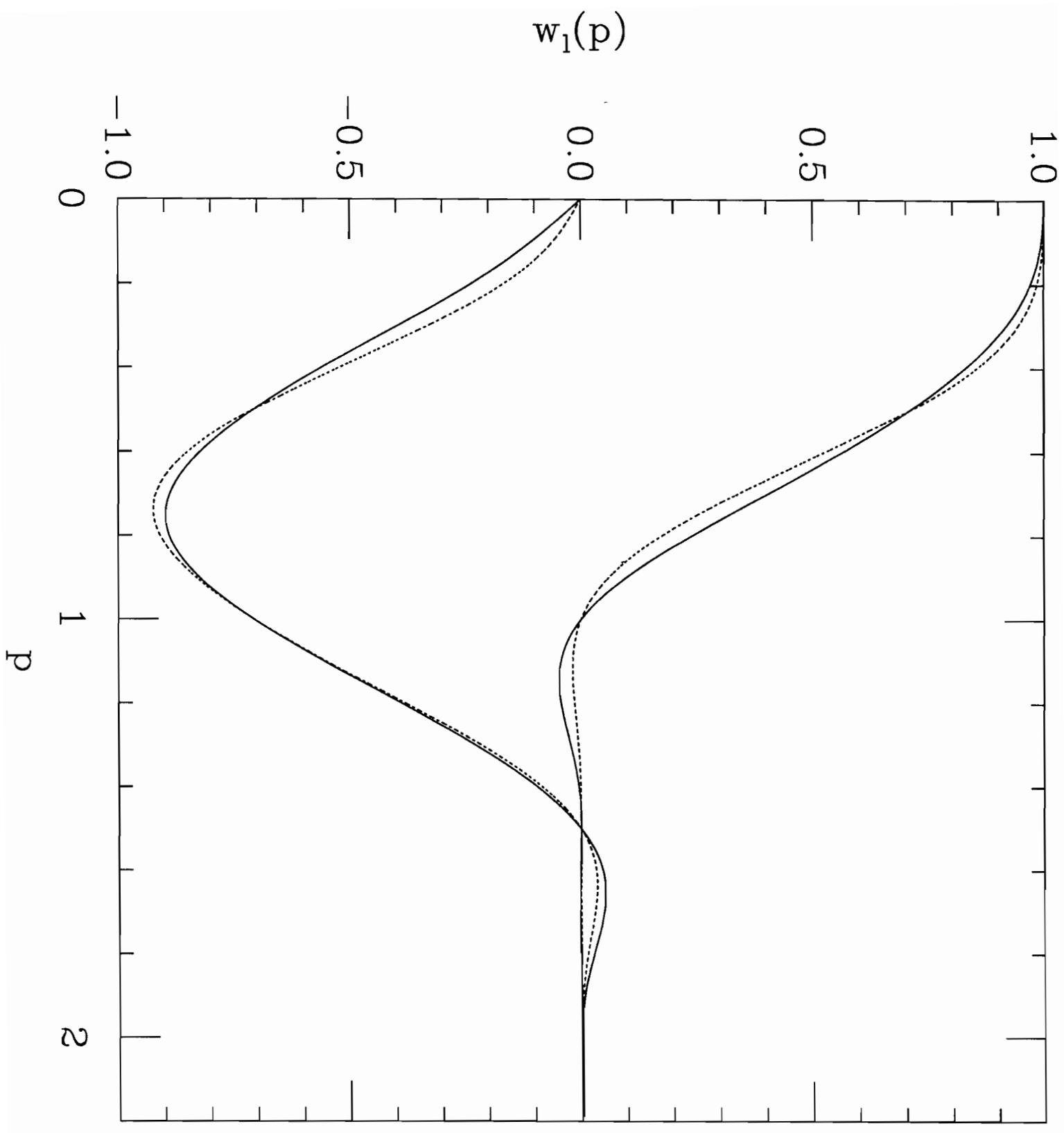
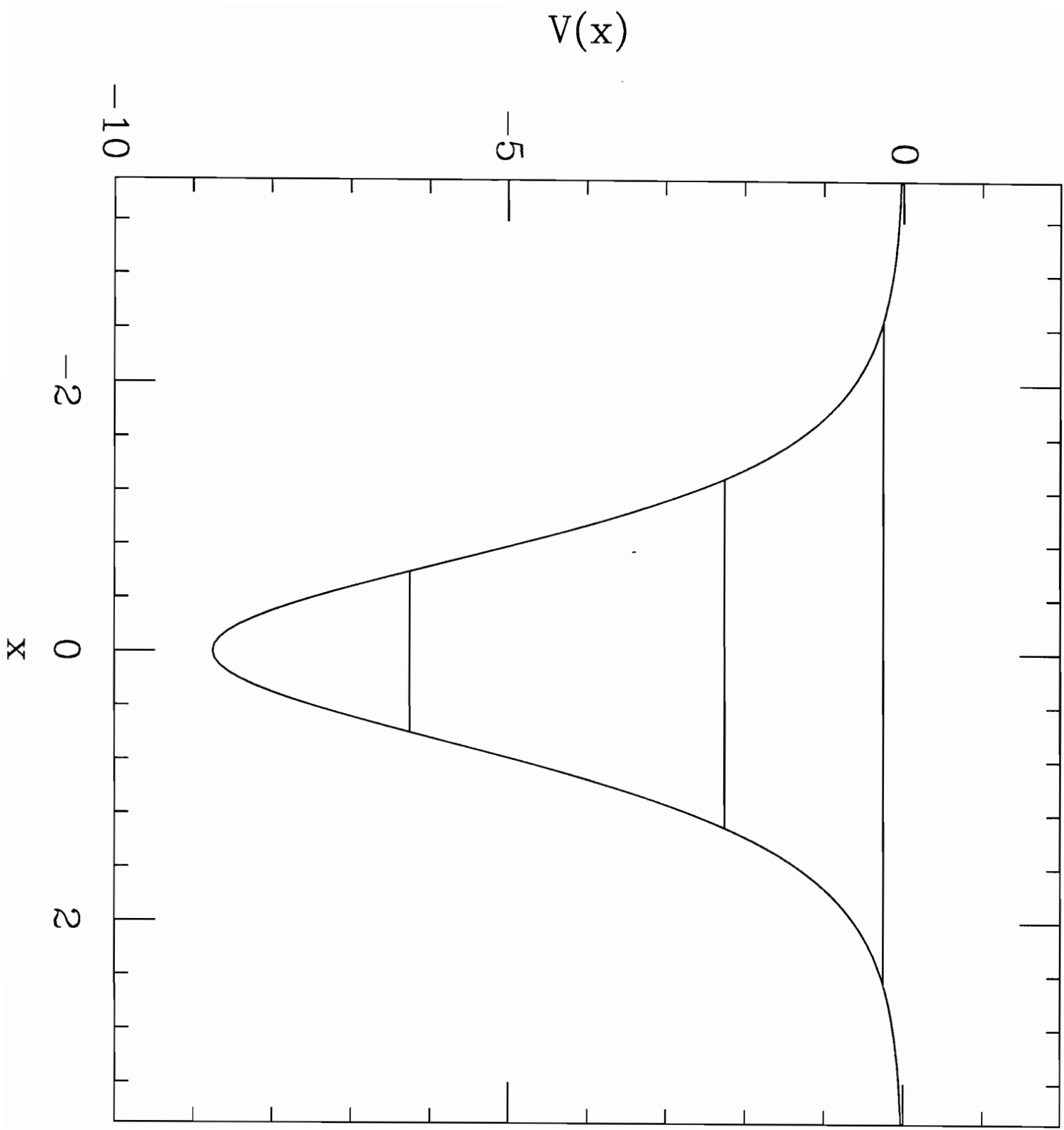
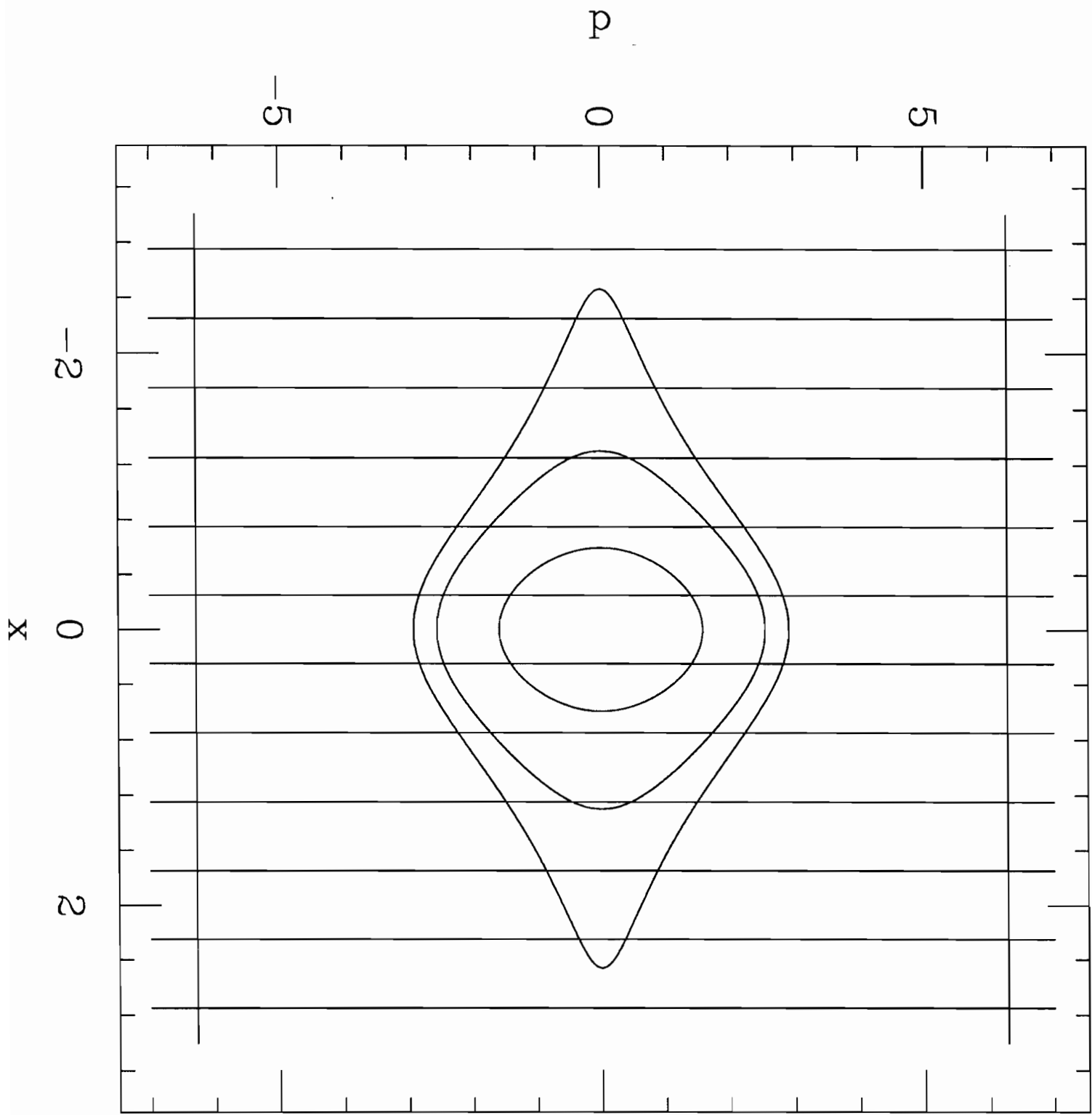
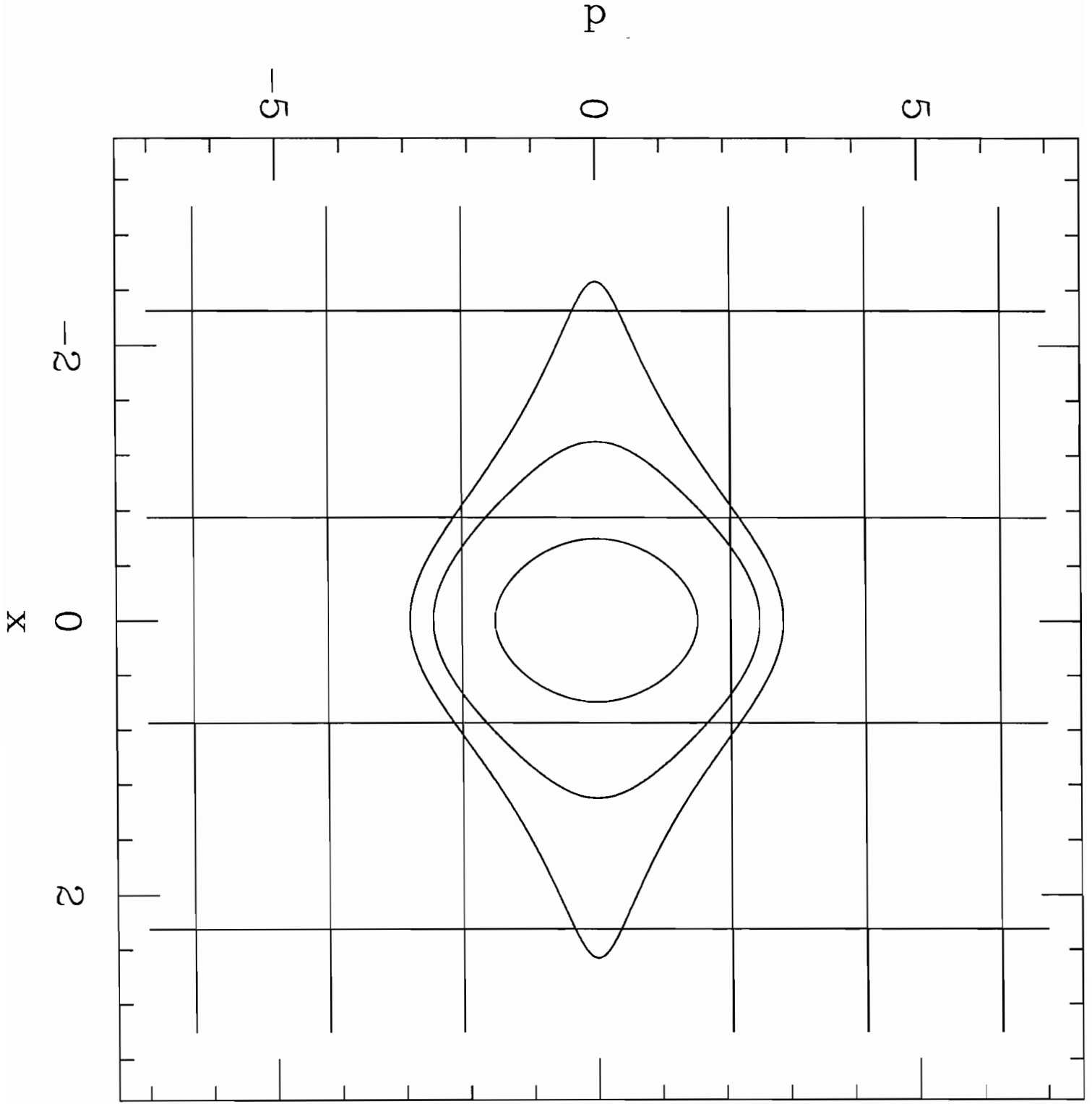
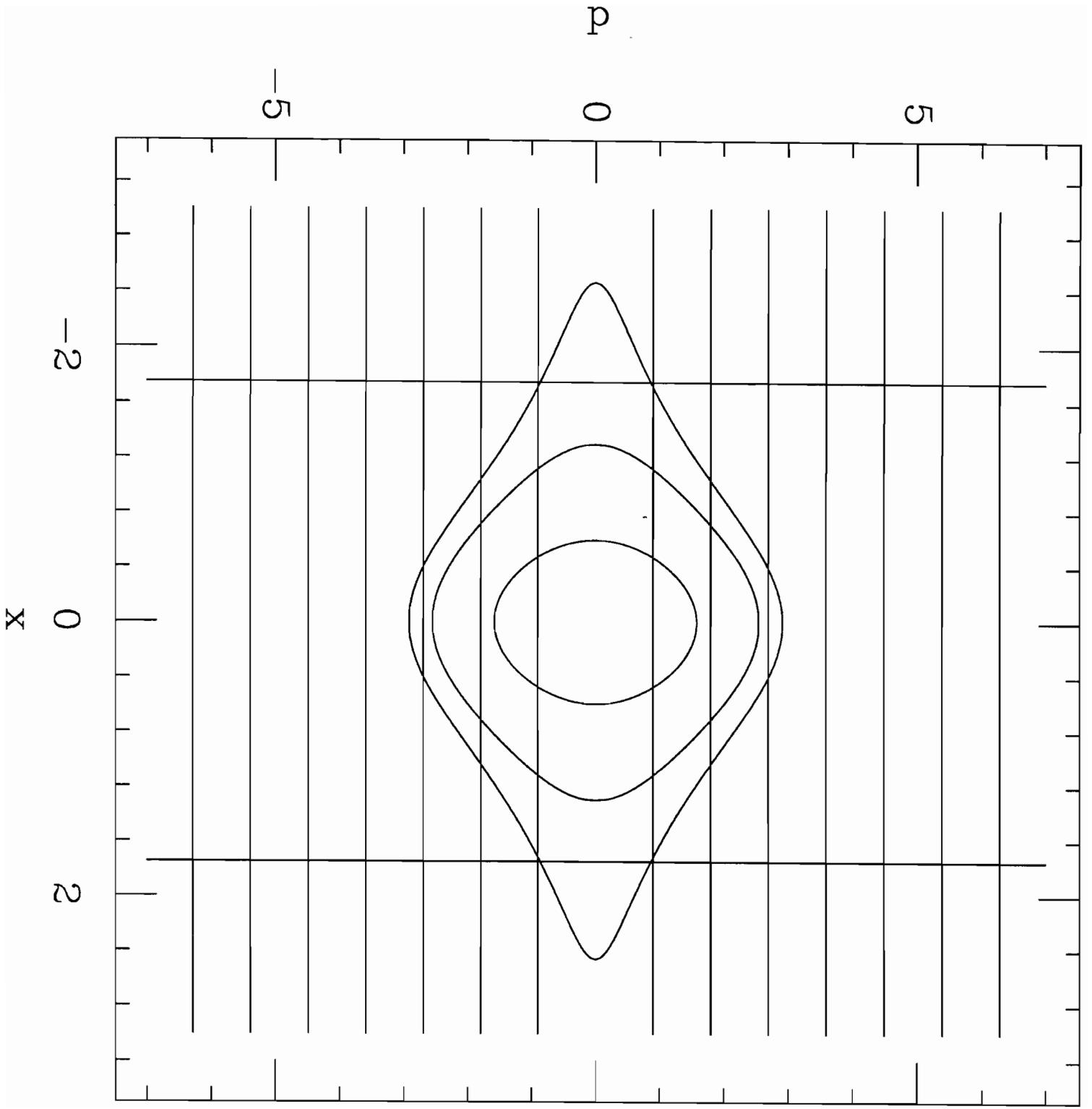
